\documentclass{aa}  
\usepackage{adjustbox}
\usepackage{caption}
\usepackage{float}
\usepackage{graphicx}
\usepackage{xcolor}
\usepackage{txfonts}
\usepackage[]{hyperref}
\hypersetup{
    colorlinks=true,
    linkcolor=blue,
    citecolor=blue,
    filecolor=magenta,      
    urlcolor=magenta,
    }
\usepackage{morefloats}
\extrafloats{100}
\usepackage{soul}
\usepackage{subcaption}

\usepackage{xurl} 
\usepackage{booktabs}
\usepackage{array}
\usepackage{orcidlink}

\newcommand{\refrep}[1]{{\textbf{ [#1]}}}

\usepackage[switch]{lineno}

\begin{document} 

   \title{ \texttt{VIPCALs}: A fully automated calibration pipeline for very long baseline interferometry data}

   \subtitle{}

   \authorrunning{D. \'Alvarez-Ortega et al.}
   \author{D. \'Alvarez-Ortega\inst{1,2}\orcidlink{0000-0002-9998-5238}
          \and
          C. Casadio\inst{1,2}\orcidlink{0000-0003-1117-2863}
          \and
          F. M. P{\"o}tzl\inst{1,2}\orcidlink{0000-0002-6579-8311}
          \and
          A. Kumar\inst{1,2}\orcidlink{0000-0002-7994-5167}
          \and
          M. Janssen\inst{3}\orcidlink{0000-0001-8685-6544}
          }
   \institute{
           Institute of Astrophysics, Foundation for Research and Technology – Hellas, N. Plastira 100, Voutes 70013, Heraklion, Greece
           \email{dalvarez@physics.uoc.gr}
       \and
           Department of Physics, University of Crete, 70013 Heraklion, Greece
       \and
           Department of Astrophysics, Institute for Mathematics, Astrophysics and Particle Physics (IMAPP), Radboud University, P.O. Box 9010, 6500 GL Nijmegen, The Netherlands
   }
  \abstract
   {Very long baseline interferometry (VLBI) is a powerful observational technique that can achieve sub-milliarcsecond resolution. However, it requires complex and often manual post-correlation calibration to correct for instrumental, geometric, and propagation-related errors. Unlike connected-element interferometers, VLBI arrays typically deliver raw visibilities rather than science-ready data, and existing pipelines are largely semi-automated and reliant on user supervision.}
   {We aim to develop and validate a fully automated, end-to-end calibration pipeline for continuum VLBI data that operates without human intervention or prior knowledge of the dataset. The pipeline must be scalable to thousands of sources and suitable for heterogeneous archival observations, as required by initiatives such as the Search for Milli-Lenses (SMILE) project.}
   {We present the VLBI Pipeline for automated data Calibration using \texttt{AIPS}, or \texttt{VIPCALs}. Implemented in Python using \texttt{ParselTongue}, \texttt{VIPCALs} reproduces the standard \texttt{AIPS} calibration workflow in a fully unsupervised mode. The pipeline carries out data import, retrieval of system temperature and gain curve data, ionospheric and geometric corrections, fringe fitting, and amplitude and bandpass calibration steps. \texttt{VIPCALs} performs automatic reference antenna selection and calibrator identification, and it generates diagnostic outputs for inspection. It can be easily used through a simple graphical user interface. We validated \texttt{VIPCALs} on a representative sample of Very Long Baseline Array (VLBA) data corresponding to 1000 sources from the SMILE project.}
   {\texttt{VIPCALs} successfully calibrated observations of 955 of the 1000 test sources across multiple frequency bands. Over 91\% of the calibrated datasets achieved successful fringe fitting on target in at least half of the solutions attempted. The median ratio of calibrated visibilities to initial total visibilities was 0.87. The average processing time was below 10 minutes per dataset when using a single-core configuration, demonstrating both efficiency and scalability.}
   {\texttt{VIPCALs} enables robust, reproducible, and fully automated calibration of VLBI continuum data, significantly lowering the entry barrier for VLBI science and making large-scale projects such as SMILE feasible.}


   \maketitle

\section{Introduction}

Very long baseline interferometry (VLBI) is a powerful radio interferometric technique that combines signals from widely separated radio telescopes to achieve up to sub-milliarcsecond angular resolution. In VLBI, each antenna independently records the incoming radio signal along with precise timing information, usually provided by a hydrogen maser clock. These signals are brought together and cross-correlated in pairs of antennas, referred to as baselines, using specialized software or hardware correlators. The result is a set of complex values, known as visibilities, that sample the Fourier transform of the source's brightness distribution in the sky. However, the measured visibilities are affected by a range of systematic effects that arise from both the different signal propagation paths and the instrumentation at each station. Although modern VLBI correlators apply corrections based on accurate physical and geometrical models, significant residual errors remain that must be addressed through post-correlation calibration.

The process of post-correlation calibration in VLBI is inherently complex and requires both domain expertise and proficiency with specialized calibration software. The most widely used packages for calibration are the Astronomical Image Processing System (\texttt{AIPS}; \citealt{Greisen2003}), the Common Astronomy Software Applications package (\texttt{CASA}; \citealt{McMullin2007}), the Haystack Observatory Post-processing System (\texttt{HOPS}; \citealt{Whitney2004}), and \texttt{PIMA} \citep{Petrov2011}. In contrast to some connected-element interferometers, such as the Karl G. Jansky Very Large Array (VLA) and the Atacama Large Millimeter Array (ALMA), where principal investigators (PIs) can request science-ready data products, VLBI projects typically only provide raw correlated visibilities to the PI. Although the delivery of calibrated data products is a goal for next-generation facilities such as the Square Kilometre Array (SKA) and the next-generation VLA (ngVLA), it is not yet the norm within the VLBI community. A partial exception is the European VLBI Network (EVN), which offers an automated calibration pipeline \citep{Reynolds2002}. However, the resulting datasets are explicitly labeled as not science ready.

Although several semi-automated VLBI calibration pipelines exist, such as \texttt{VLBARUN} in \texttt{AIPS} and \texttt{rPICARD} in \texttt{CASA} \citep{Janssen2019}, none operate in a fully unsupervised manner. These algorithms typically require user interaction to identify calibrator sources, validate fringe-fitting results, or adjust calibration parameters. Such manual oversight becomes a critical bottleneck for large-scale VLBI surveys involving thousands of radio sources. This is the case for the Search for Milli-Lenses (SMILE) project \citep{Casadio2021}, which aims to identify strong gravitational lensing systems at milliarcsecond angular scales using archival Very Long Baseline Array (VLBA) observations. The SMILE sample consists of nearly 5000 radio-loud sources, observed over multiple decades with diverse frequency setups and observing strategies. This level of heterogeneity renders the use of semi-automated pipelines impractical.

To address these limitations, we have developed the VLBI
Pipeline for automated data Calibration using \texttt{AIPS} (in the
SMILE framework), or \texttt{VIPCALs}. Although originally motivated by the requirements of the SMILE project, the pipeline is designed as a general-purpose tool capable of unsupervised calibration across a wide range of continuum VLBI datasets. It operates end-to-end without manual intervention or prior information on the data, from the initial data loading to the generation of science-ready visibility products. To validate its performance, the pipeline has been applied to a representative sample of VLBA observations of 1000 different sources encompassing a wide range of observational configurations and signal-to-noise conditions. This paper describes the pipeline’s implementation, its automated calibration workflow, and its performance on this large and heterogeneous dataset.

In Section \ref{sec:vlbicalib} we introduce the data calibration process in VLBI and more specifically in \texttt{AIPS}. Section \ref{sec:workflow} provides a thorough description of the calibration workflow followed by \texttt{VIPCALs}. In Section \ref{sec:sample} we describe the VLBA sample used for validation, while the results and their discussion are presented in Section \ref{sec:results}. Section \ref{sec:futurework} gives an overview on future development, and finally, a summary and conclusions are presented in Section \ref{sec:conclusions}.

\section{VLBI data calibration}\label{sec:vlbicalib}
The objective of VLBI data calibration is to correct for instrumental, geometric, and propagation-induced effects that affect both the phase and amplitude of the observed complex visibilities. These corrections are essential for coherent averaging of the signals of different antennas and ultimately to recovering accurate astronomical data. Phase calibration focuses on correcting residual phase errors that remain after initial geometric modeling during correlation. These residuals manifest primarily as group delays and fringe rates, corresponding to the derivatives of phase with respect to frequency (group delay) and time (fringe rate). Such errors arise from several sources, including mismodeling of the geometric delay due to imperfect knowledge of antenna positions or Earth orientation parameters, unmodeled atmospheric propagation effects, and clock or local oscillator instabilities at each station, among others. The final goal is to produce phase-stable data over time and frequency, thereby enabling coherent averaging and enhancement of the signal-to-noise ratio (S/N). On the other hand, amplitude calibration is performed to correct for variations in system gain, system temperature, and instrumental bandpass response on each antenna. These corrections ensure that the visibilities are on a consistent flux scale and can be reliably interpreted in subsequent analysis steps.

For many decades, \texttt{AIPS}, developed by the National Radio Astronomy Observatory (NRAO), has served as the primary software environment for the calibration and reduction of radio interferometric data. \texttt{AIPS} provides a full calibration framework, with dedicated tools to correct for instrumental and propagation effects, manage metadata, inspect data quality, and prepare calibrated datasets for imaging and analysis. Due to its long-term development, community support, and robust documentation, \texttt{AIPS} remains a cornerstone of VLBI data reduction workflows and is still widely used alongside more modern packages.

\subsection{\texttt{AIPS} table scheme}
In \texttt{AIPS}, the raw interferometric data are stored as visibility datasets, which contain the measured complex visibilities for each baseline as a function of time, frequency, and polarization. The \texttt{AIPS} calibration framework, as well as that of other calibration software such as \texttt{CASA} or \texttt{PIMA}, is based on the Hamaker-Bregman-Sault measurement equation \citep{Hamaker1996}. This equation provides a matrix formalism to describe the full polarization response of an interferometer. For a pair of antennas, $i$ and $j$, the observed visibility matrix, $\mathcal{V}^{\text{obs}}_{ij}$, is given by

\begin{equation}
    \mathcal{V}_{ij}^{\text{obs}} = J_i \;\mathcal{V}_{ij}^{\text{true}}\; J_j^{\dagger},
    \label{equation:measeq}
\end{equation}

where $\mathcal{V}^{\text{true}}_{ij}$ is the true sky coherence matrix, and $J_i$ and $J_j$ are the Jones matrices describing the instrumental and propagation effects of each antenna, where $\dagger$ denotes the Hermitian conjugate.

The calibration, in essence, consists of determining these Jones matrices as accurately as possible in order to invert Equation~\ref{equation:measeq} and recover the true visibilities from the observed measurements. In \texttt{AIPS}, this process is carried out through tasks, i.e., standalone programs designed to create, manipulate, and apply calibration tables to the data. In the following sections, \texttt{AIPS} task names will appear in uppercase italic, while their different adjustable parameters will be indicated in lowercase italic for clarity.

Alongside the visibility data, \texttt{AIPS} maintains a set of auxiliary files known as extension tables, which contain metadata and calibration information essential for the processing. An incremental calibration strategy is adopted, wherein calibration corrections are applied in stages and stored in the so-called calibration (CL) tables. These tables contain amplitude and phase corrections for each antenna, representing the cumulative effect of various calibration steps. Each time a new calibration is applied, a new version of the CL table is generated, preserving the previous calibration states and essentially encoding the Jones matrices per antenna. These calibration solutions are not applied to the visibility data until the data are displayed or exported.

Intermediate calibration solutions are stored in solution (SN) tables. These tables serve as a staging area for newly derived corrections, such as those from fringe fitting or the digital sampling correction, before they are interpolated and applied to the CL tables. This separation allows for the inspection and editing of individual solution steps prior to their incorporation into the final calibration.

The bandpass (BP) table contains the complex frequency-dependent gain response of each antenna, which models instrumental distortions across the observing bandwidth. It effectively constitutes another Jones matrix, and it is applied in a similar way as the CL tables. Something similar occurs in  polarization calibration, where the polarization D-term (PD) table stores estimates of instrumental polarization leakage, derived from observations of unpolarized or well-characterized calibrators.

Other important tables with relevant information include the following:
\begin{itemize}
    \item Index table (NX): Defines time boundaries, sources, subarrays, and location within the file of each scan within an observation.
    \item Antenna table (AN): Contains antenna names and coordinates.
    \item Source table (SU): Lists observed sources, their positions, and their fluxes if available.
    \item System temperature table (TY): Provides per-antenna system temperature measurements as a function of time and frequency.
    \item Gain curve table (GC): Contains the expected zenith gain and gain-elevation curve for each antenna.
    \item Weather table (WX): Contains weather-related information for each station, including opacity measurements when available.
    \item Flag table (FG): Stores data flagging information.
\end{itemize}

\subsection{Fringe fitting in \texttt{AIPS}}\label{sec:fringefitAIPS}

Fringe fitting is the primary method for phase calibration in VLBI data reduction. It solves for the residual phase, delay (phase versus frequency), and fringe rate (phase versus time) between each antenna and a designated reference antenna, thereby compensating for any unmodeled effects. The objective is to align the signals from different stations in time and frequency, maximizing the coherence of the signal.

In \texttt{AIPS}, fringe fitting is implemented through the task \textit{FRING}, which performs a form of global fringe fitting. This approach, based on the algorithm by \cite{Schwab1983}, simultaneously solves for antenna-based delay and rate solutions across all baselines, rather than treating each baseline independently, hence the term global. It operates in two main steps: a coarse search followed by a least-squares refinement. First, a two-dimensional fast Fourier transform (FFT) is applied to the complex visibilities over time and frequency, transforming the data into the delay-rate space. In this space, the algorithm searches for the peak of the fringe amplitude. For that purpose, the fringe S/N is defined as the amplitude of the peak in the delay-rate plane divided by the root-mean-square of the surrounding noise. The initial search is conducted on a coarse grid whose size is tunable, and only peaks exceeding a specified S/N threshold are considered valid fringe detections. This fringe S/N not only determines the reliability of the fringe detection, but also serves as a useful metric throughout the calibration. In the second step, delay and rate estimates for antennas with valid fringe detections are refined using a global least-squares solver. This solver assumes a point-source model at the phase center of the scan, and adjusts all antenna solutions simultaneously. In cases where this assumption may not hold (e.g., resolved or complex sources), fringe-fitting performance can be improved by supplying a more accurate source model.

Since phase is inherently a relative quantity, fringe solutions must be referenced to a specific antenna. The delay and rate for the reference antenna are fixed to zero, and all other antenna-based solutions are determined relative to it. By default, \texttt{AIPS} performs the FFT search between each antenna and the reference antenna. If the algorithm fails to find a valid solution for a given antenna with respect to the reference, it attempts to fringe fit that antenna against at least one alternative antenna. If specified by the user via the \textit{aparm(9)} parameter, all antennas can be searched for as alternative reference antennas. A prioritized order can be given through the \textit{search} parameter. If this succeeds, then the delay and rate solutions are re-referenced to the primary reference antenna.

\subsection{The \texttt{VIPCALs} pipeline}

The calibration of VLBI data remains a highly manual and expertise-driven process. Each stage typically requires careful inspection of intermediate solutions and manual decision-making, such as selecting the reference antenna, choosing appropriate calibrator sources, and fine-tuning parameters for fringe fitting and other calibration tasks. This process can be time consuming, often requiring significant human oversight. As a result, it becomes particularly challenging to scale to large projects including hundreds or thousands of sources, especially when they involve heterogeneous observations. In such cases, manual calibration can present a major bottleneck in both time and resources. This is accentuated by the fact that, although \texttt{AIPS} provides a comprehensive suite of tools for VLBI data reduction, its learning curve can also pose a significant barrier to new users. Manual calibration also hinders reproducibility, since decisions made interactively are often difficult to document and repeat exactly.

To address these challenges, particularly those arising in the context of large VLBI surveys, we have developed \texttt{VIPCALs}.\footnote{\url{https://github.com/dalvarezo/VIPCALs}} This automated pipeline was motivated by the needs of the SMILE project \citep{Casadio2021}, which aims to identify gravitational lens candidates on milliarcsecond scales (milli-lenses) by analyzing VLBI datasets of approximately 5000 radio-loud galaxies. The abundance of such milli-lens systems are a key discriminant between dark matter models \citep{Loudas2022}. As the project primarily relies on archival cm-wavelength VLBA data, the volume and diversity of datasets make manual calibration and imaging impractical. While a dedicated automatic imaging solution is currently under development (P\"otzl et al., in prep.), \texttt{VIPCALs} provides a fully automated pipeline for calibration, requiring no prior knowledge of the datasets and delivering science-ready visibility data products.

The pipeline has been implemented using \texttt{ParselTongue} \citep{Kettenis2006}, a Python interface to \texttt{AIPS}, which enables flexible scripting and integration with widely used Python packages such as \texttt{matplotlib} for diagnostics and plotting. \texttt{ParselTongue} provides full access to calibration tables and visibility data, as well as the ability to execute \texttt{AIPS} tasks programmatically, making it an ideal framework for developing automated data reduction pipelines. To improve usability, \texttt{VIPCALs} implements a simple graphical user interface (GUI) built with PySide6, allowing users with limited or no \texttt{AIPS} experience to interact easily with the pipeline. \texttt{VIPCALs} is available both as a pip-installable package (requiring a dedicated Conda environment and a local installation of \texttt{AIPS} \texttt{31DEC24} or newer) and as a self-contained Docker container.

Running the pipeline requires only minimal input from the user: the visibility data in \texttt{uvfits} or \texttt{idifits} format, an output directory, and the names of the sources to be calibrated (or an option to calibrate all). When used with a local \texttt{AIPS} installation, the user must also specify the \texttt{AIPS} disk and user number. From these inputs, the pipeline performs a complete calibration of the dataset, producing fully calibrated visibility files along with a series of diagnostic plots and statistical summaries, as described in Section~\ref{sec:outputs}. In addition to the native \texttt{AIPS} logs, a structured and human-readable log is generated, detailing each calibration step and its results. While designed for full automation, the pipeline also supports fine-tuning of numerous calibration parameters, enabling more experienced users to guide the calibration process when prior knowledge of the sources or observing conditions is available.

\begin{figure*}[h]
    \centering
    \includegraphics[width=0.95\textwidth]{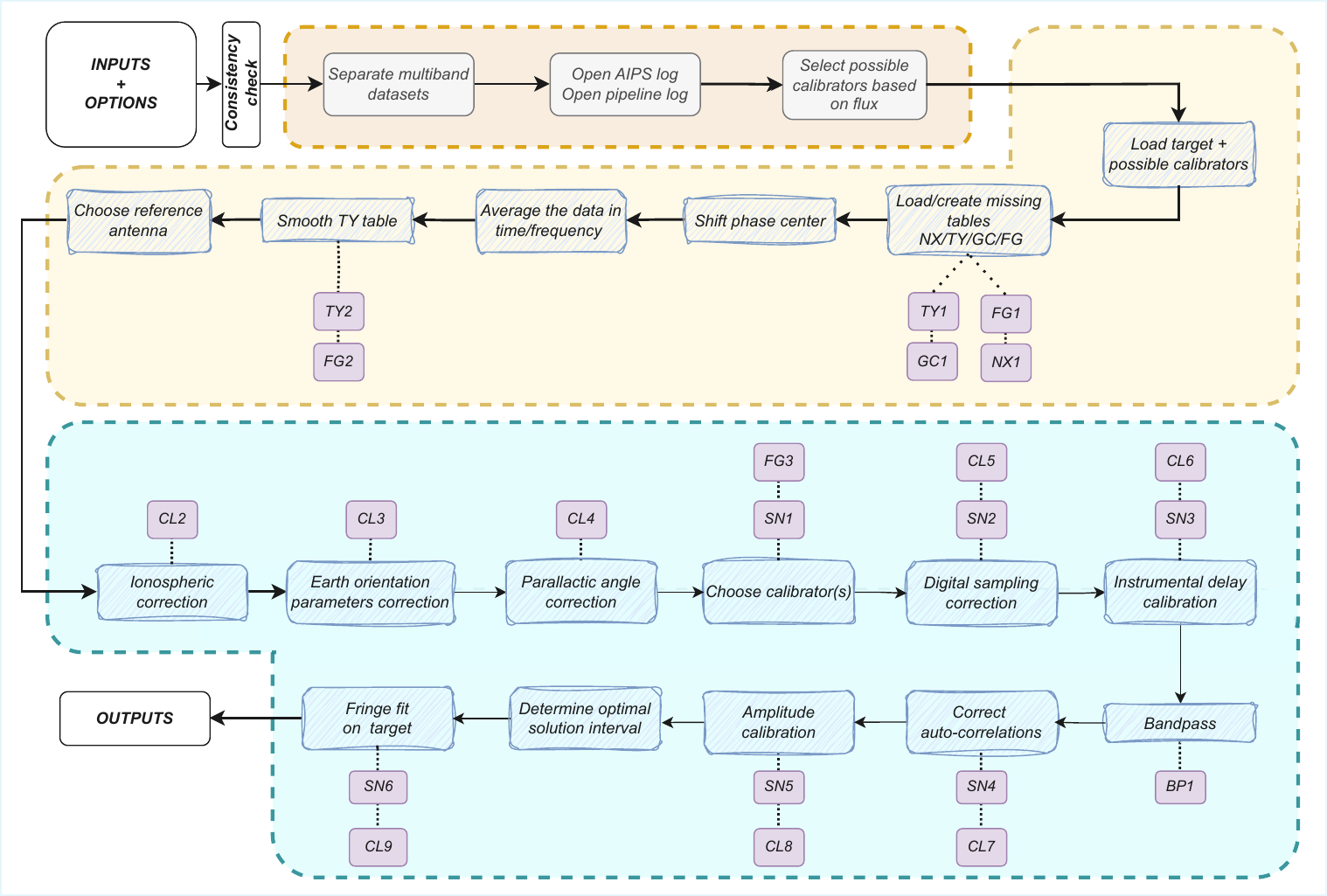}
    \caption{Overview of the \texttt{VIPCALs} calibration pipeline. The orange block represents the preloading stage, the yellow block corresponds to loading and precalibration tasks, and the blue block encompasses the phase and amplitude calibration steps. Purple boxes indicate the \texttt{AIPS} tables generated at each stage of the workflow.}
    \label{fig:workflow}
\end{figure*}

\section{Calibration workflow}\label{sec:workflow}
In the following section, we provide a detailed description of each calibration step performed by the pipeline, as summarized in Figure~\ref{fig:workflow}. The strategy follows the standard procedures from the \texttt{AIPS} Cookbook,\footnote{\url{https://www.aips.nrao.edu/cook.html}} and a summary of all the different \texttt{AIPS} tables used can be seen in Table~\ref{table:tables}. Given that the primary goal of the pipeline is to support the SMILE project, the process has been optimized for centimeter-wavelength, continuum VLBI observations of radio-loud active galactic nuclei with flux densities exceeding 10~mJy. Despite this, the approach should be suitable for most continuum datasets, with some limitations noted in the text (see Section~\ref{sec:faileddata}). An important consideration is that the calibration of polarization observations is currently not supported, as automating the process, especially the correction of instrumental polarization (D-terms), remains a significant challenge. This functionality is left for future development.

\subsection{Data loading}
The pipeline begins by checking whether the dataset contains observations at widely separated central frequencies, either through distinct intermediate frequencies (IFs) or different frequency identifiers (IDs). If such cases are detected, the data are imported into \texttt{AIPS} as separate entries to simplify the calibration process. A typical example is the use of the VLBA’s 13/4-cm dichroic system, which records data simultaneously at 13 cm (S-band) and 4 cm (X-band). In such observations, some IFs correspond to the S-band and others to the X-band. The pipeline also supports the concatenation of multiple \texttt{uvfits} or \texttt{idifits} files, as long as their frequency setups are identical in terms of central frequencies, bandwidths, and channel configurations.

To optimize performance, the pipeline selectively loads only the science targets of interest along with up to three bright calibrator candidates. This is particularly important for SMILE datasets, which often contain hundreds of unrelated sources that are not relevant to the analysis. Excluding these sources significantly reduces the processing time. Potential calibrators are identified without any prior metadata by cross-matching the coordinates of observed sources against the NRAO’s VLBA calibrator list,\footnote{\url{http://www.vlba.nrao.edu/astro/calib/vlbaCalib_allfreq_full.txt}} which contains positions and flux densities across 2–23\,GHz for over 9000 sources. A conservative matching radius of 1 arcsecond is used to minimize false cross-identifications. The three brightest matched sources in the observed band are loaded as potential instrumental calibrators. If the observing band is not included in the calibrator list, or if no matches are found, all sources in the dataset are loaded to ensure that no potential calibrator is excluded. Users may override this behavior by enabling from the beginning the \textit{load all} option on the pipeline.

\begin{table*}[t]
\caption{Summary of \texttt{AIPS} tables produced by \texttt{VIPCALs} during the different calibration stages.}
\label{table:tables}
\renewcommand{\arraystretch}{1.2}
\centering
\begin{tabular}{@{}ll@{\hspace{2cm}}ll@{}}
\toprule
\textbf{Table} & \textbf{Content} & \textbf{Table} & \textbf{Content} \\
\midrule
CL1  & Initial calibration table                         &  SN1  & Fringe S/N of all sources  \\
CL2  & Applies ionospheric corrections                   &  SN2  & Normalized auto-correlations   \\
CL3  & Applies EOP corrections                           &  SN3  & Delay corrections of calibrator scan(s) \\
CL4  & Applies parallactic angle corrections             &  SN4  & Normalized auto-correlations (after bandpass) \\
CL5  & Applies digital sampling corrections              &  SN5  & Amplitude scaling  \\
CL6  & Applies instrumental delay corrections            &  SN6  & Residual phase, delay, and rate (science target \\
CL7  & Applies normalization of auto-correlations &    &  or phase reference calibrator)  \\
     & after bandpass                             &    & \rule{6cm}{0.4pt}                                   \\
CL8  & Applies amplitude calibration                     & BP1  & Bandpass corrections   \\
CL9  & Applies fringe-fitting solutions to science target(s)    &  & \rule{6cm}{0.4pt}  \\
     & \rule{6cm}{0.4pt}                                 & TY1   & Initial system temperatures \\
FG1    & Initial flag table                               & TY2   & Filtered system temperatures \\
FG2    & Flags antennas missing TY/GC                 &  GC1   & Antenna gain curves \\
FG3    & Flags antennas with no fringe detections    & NX1   & Index table\\
       & over the S/N threshold                           &      &                \\
\bottomrule
\end{tabular}
\end{table*}

Data are loaded through the \textit{FITLD} task in \texttt{AIPS}. At this stage, the pipeline also sets the entry interval for the calibration tables, i.e., the temporal spacing at which calibration solutions are stored and applied. This interval, distinct from the dataset's native time sampling, must be short compared with the expected coherence time. If it is too short, calibration tables become large and computationally inefficient; if too long, temporal variations in the instrumental or atmospheric conditions may be undersampled, degrading the calibration. Based on experience, a 6-second interval provides a good compromise between processing speed and the fidelity of the applied corrections at most wavelengths.

\subsection{Precalibration steps}
Prior to applying any calibration, a set of preparation steps are performed to ensure that the dataset is suitable for processing. Metadata fields such as source names, antenna labels, and polarization descriptors are cleaned to remove padding or formatting inconsistencies that could present any conflicts later on. The visibility data contained in the files must be ordered in time–baseline format, a requirement that should be correctly reflected in the file header. If this is not the case, the \texttt{AIPS} task \textit{UVSRT} is used to reorder the data accordingly. In case the index (NX1) table is missing, the task \textit{INDXR} generates it along the initial calibration table (CL1).

It is worth noting that, aside from a conservative smoothing of the system temperatures and the flagging information provided by the correlator, no additional a priori flagging is applied. This design choice reflects the pipeline’s conservative philosophy: to retain as much data as possible while still producing a robust and high-quality calibration. Further flagging is deferred to later stages, such as the imaging.

\subsubsection{Table retrieval}

Modern VLBI datasets typically include any necessary auxiliary tables attached to the visibility data. Among the most critical are the initial flag table (FG1), the system temperature table (TY1), the antenna gain curve table (GC1), and weather data (WX1). The TY1 and GC1 tables are especially important, as they are required for amplitude calibration. The WX1 table is used primarily for opacity corrections. When these tables are not included in the dataset, e.g., in older archival data, they must be appended manually.

For VLBA observations, FG, TY, and WX tables of each individual project are publicly available in an online repository.\footnote{\url{http://www.vlba.nrao.edu/astro/VOBS/astronomy/}}  \texttt{VIPCALs} can automatically query, retrieve, and apply minor edits to these tables to make them compatible with \texttt{AIPS}, where they are imported using the \texttt{ANTAB} task. 
As part of the preprocessing, TY tables are further refined using the \texttt{AIPS} task \textit{TYSMO}, which performs basic smoothing and data cleaning. The pipeline adopts a conservative approach, rejecting system temperature values that are negative, exceed 1500K, or deviate by more than 250K from the mean value for a given source. In contrast, GC tables are not provided on a per-project basis but are instead stored in a common gain curve file,\footnote{\url{http://www.vlba.nrao.edu/astro/VOBS/astronomy/vlba_gains.key}} which compiles periodic measurements across all VLBA observing bands. The pipeline parses this file to extract the appropriate gain curves corresponding to the antennas and time range of the observation, and loads them into \texttt{AIPS} accordingly.

It is important to note that the current version of \texttt{VIPCALs} does not support extensive table editing or modifications that require prior knowledge of the observation, such as specifying the observing mode of the Very Large Array when included (as a phased array - Y27, or with a single antenna - Y1). Additionally, calibration tables for non-VLBA antennas are not automatically retrieved by the pipeline, and as a consequence such antennas are flagged (FG2). Both limitations are currently under active development. If needed, the user can always give the necessary tables as an optional input through the pipeline interface.

\subsubsection{Phase center shift}
In the absence of a source model, fringe fitting assumes that the target is a point source located at the phase center used during correlation. If the actual source coordinates differ from this assumed position, residual phase and rate errors can be introduced. When more accurate source coordinates are available, \texttt{VIPCALs} allows for correction via the \texttt{AIPS} task \textit{UVFIX}, which uses a geometric model to apply a shift to the visibilities. 

This correction is particularly important in the context of the SMILE project, which processes a wide range of archival VLBI observations. Many of these datasets were correlated using position estimates, particularly in projects aimed at detecting sources in wide fields. To optimize fringe-fitting performance and minimize the effects of bandwidth and time smearing \citep{Bridle1989} in subsequent calibration steps, accurate phase center alignment is crucial. Based on empirical results, calibration of sources offset by more than $\sim$1 arcsecond from the correlation center shows significant improvement when this correction is applied.

\subsubsection{Time and frequency averaging}

Given the heterogeneous nature of the SMILE sample, the pipeline includes an optional precalibration averaging step in both time and frequency. This step enhances the S/N of the visibilities while keeping time and bandwidth smearing within acceptable limits for the expected field of view. By default, frequency averaging is applied when the individual channel width is below 500~kHz. Time averaging is applied only when the original integration time is 1~second or shorter, in which case the data are averaged into 2-second bins. This ensures at least two visibilities per bin and reduces the risk of flagging due to undersampling. These thresholds are based on empirical testing and reflect a balance between improved sensitivity and the need to preserve coherence of the data. Using these values in the EVN calculator,\footnote{\url{https://services.jive.eu/evn-calculator/cgi-bin/EVNcalc.pl}} the field of view of a modern VLBA observation at X-band would be limited by bandwidth smearing to 11 arcseconds, while time smearing would limit it to 7.4 arcseconds. These limits are computed as a 10\% loss in response from a point source located at the phase center. In the context of the SMILE project, where the goal is to identify potential lensed images within a field of view of approximately 300 milliarcseconds, the selected time and frequency averaging parameters result in negligible smearing effects at the relevant angular scales.

\subsubsection{Choice of reference antenna}\label{sec:refant}

The choice of reference antenna is a critical aspect of VLBI calibration. Since interferometric phases (and therefore delays and rates) are inherently relative quantities, a reference frame must be established. This is typically achieved by selecting a reference antenna whose phase, delay, and rate are set to zero, and expressing all other antenna solutions relative to it. Ideally, the reference antenna should be present throughout the entire observation, centrally located within the array to ensure uniform baseline coverage, and should possess good sensitivity, as it will be used for the initial fringe search for all baselines.

In standard practice, the reference antenna is chosen manually based on these criteria. If no single antenna fulfills all conditions, multiple candidates may be selected, as \texttt{AIPS} can re-reference solutions a posteriori.

In the \texttt{VIPCALs} pipeline, the reference antenna is selected automatically based on a set of prioritized criteria. It follows this hierarchy:

\begin{itemize}
\item Antennas must appear in all scans, including both calibrator and target observations; those that do not are excluded.
\item If no antenna satisfies the first condition, only antennas present in all science target scans are retained.
\item Optionally, for VLBA observations, preference is given to antennas located near the geographic center of the array (KP, LA, PT, OV, FD); others are discarded if any central-array candidates remain.
\item From the remaining antennas, their priority as reference antenna is determined by their average fringe S/N  across all baselines and sources.
\end{itemize}

The way the fringe S/N is computed is by running the FFT step of the fringe fit using the \textit{FRING} task. It solves for the single band delay and rate of all scans, using each antenna as the reference antenna. Solutions are searched for in a predefined window of 1000 ns and 200 mHz following prescriptions in the \texttt{AIPS} Cookbook. The fringe S/N threshold is removed setting \textit{aparm(7)} = 1 in order to prevent the task from using alternative antennas in the search. The solution interval for the fringe fit solutions is defined as the scan length. Then, the average fringe S/N of each antenna $i$ on a certain scan is computed as

\begin{equation}
    \overline{\text{S/N}}(i) = \frac{1}{(N_{\text{ant}}-1)N_{\text{IF}}N_{\text{pol}}}\sum_{\substack{j \\j \ne i}}^{N_{\text{ant}}}\sum_{f=1}^{N_{\text{IF}}}\sum_{p=1}^{N_{\text{pol}}}\text{S/N}(ij, f, p)
\end{equation}

where $N_{\text{IF}}, N_{\text{pol}},$ and $N_{\text{ant}}$ are the number of IFs, polarizations, and antennas, respectively. $\text{S/N}(ij, f, p)$ is the fringe S/N obtained at the baseline between antennas $i$ and $j$, on IF $f$, and polarization $p$. By default and in order to speed up this step, \texttt{VIPCALs} limits the search to ten randomly selected scans per source.

This procedure yields a prioritized list of candidate reference antennas. The highest-ranking antenna is used as the primary reference antennas, while the remaining candidates are passed to the \textit{search} parameter in the \textit{FRING} task. As described in Section~\ref{sec:fringefitAIPS}, these serve as fallbacks for fringe detection in case the primary antenna fails to find valid solutions.

\subsection{Amplitude and phase calibration}
The following subsections describe the corrections applied to mitigate atmospheric and instrumental effects on the visibility amplitudes and phases.

\subsubsection{Ionospheric delay calibration}

The Earth's ionosphere, composed of a magnetized plasma of free electrons and ions, introduces frequency-dependent propagation effects on radio waves. One of the most relevant effects for VLBI observations is a dispersive group delay, caused by the reduced refractive index of the ionospheric plasma. This delay results in an excess path length, particularly significant at lower observing frequencies, that depends on the observing frequency $\nu$, the zenith angle $z$, and the total electron content (TEC) along the line of sight as

\begin{equation}
\tau_{\text{iono}} \simeq \frac{40.3}{\nu^2} \cdot \sec z \cdot \int N_e ds,
\end{equation}

where $\tau_{\text{iono}}$ is the ionospheric group delay in seconds, $\nu$ is the observing frequency in Hz, and the integral represents the TEC in electrons per square meter along the ray path.

In addition to delay, the ionosphere also induces Faraday rotation, a rotation of the polarization plane of the wave due to the presence of the geomagnetic field. This effect is also dispersive and becomes relevant when calibrating full-polarization data.

With the task \textit{TECOR}, \texttt{AIPS} can correct for both ionospheric dispersive delay and Faraday rotation. \textit{TECOR} applies corrections based on global TEC maps provided in the IONospheric map EXchange format (IONEX). \texttt{VIPCALs} automates this step by retrieving the appropriate IONEX files corresponding to the observation dates, and handling any changes in file naming conventions across different epochs.\footnote{\url{https://www.earthdata.nasa.gov/data/space-geodesy-techniques/gnss/atmospheric-products}} The ionospheric corrections are applied to calibration table CL2. For observations prior to June 1998, before the availability of IONEX maps, the initial calibration table CL1 is simply copied to CL2, as no TEC-based correction can be applied.

\subsubsection{Geometric effects}
Accurate VLBI calibration requires correcting for geometric effects arising from Earth's motion relative to the celestial reference frame. These include precession, nutation, polar motion, and changes in the length of day, collectively described by the Earth Orientation Parameters (EOPs). Although predicted EOPs are typically used during correlation, applying a posteriori corrections with observed values improves positional accuracy. In \texttt{AIPS}, these corrections are applied using the task \textit{CLCOR}, which updates the calibration table by interpolating the EOP values from external files\footnote{\url{https://www.earthdata.nasa.gov/centers/cddis-daac/iers-rapid-service-prediction-center-earth-orientation-parameters}} provided by the United States Naval Observatory (USNO). These corrections are applied into the CL3 calibration table. This step can only be applied to datasets correlated with DiFX or the legacy VLBA correlator, as they include the parameters of the delay model used by the correlator. The EOPS correction is skipped for other types of datasets (e.g., for EVN data).

Additionally, geometric effects include also the parallactic angle rotation, which arises due to the apparent rotation of the sky relative to the fixed orientation of each antenna's feed. While the impact is most pronounced in full-polarization observations, where it leads to a rotation over time of the observed polarization angle in the antenna frame, it can also influence total intensity data under certain conditions. As such, correcting for parallactic angle rotation is a standard step in the calibration process. Although \texttt{VIPCALs} does not yet implement full polarization calibration, the correction for parallactic angle is already included as a preparatory step. This correction is also performed using \textit{CLCOR} and is stored in the CL4 calibration table.

\subsubsection{Choice of calibrator scans}\label{sec:calibchoice}

Very long baseline interferometry observations require strong calibrator sources to correct for instrumental effects, particularly instrumental delays and bandpass variations. Ideally, these calibrators should be bright, unresolved, and well distributed throughout the observation to ensure reliable calibration across the full dataset. However, due to scheduling constraints, target prioritization, or observational limitations, such ideal conditions are not always met.
To identify suitable calibrator scans, \texttt{VIPCALs} performs a fringe fit on all scans in the dataset, solving for single-band delay and fringe rate and stopping at the FFT stage. The results are stored in the solution table SN1, which contains fringe S/Ns for each scan and baseline to the reference antenna. The parameters are the same as in \ref{sec:refant}, with the addition of fringe S/N threshold of 5 to ensure that only scans with significant detections are taken into account. Based on these results, the pipeline selects, for each antenna, the scan with the highest fringe S/N and defines this as its calibrator scan. Ideally, all antennas would identify the same calibrator scan; however, in practice, this varies across the array. This method prioritizes signal strength but does not guarantee that the chosen sources are unresolved or temporally adjacent to the science targets.

If the highest fringe S/N for a given antenna does not exceed the threshold of fringe S/N of 5, that antenna is flagged and excluded from further calibration. Since the fringe fit is performed with the search parameter, failing to meet this threshold indicates that no sources were reliable detected with any baseline to that antenna. Antennas flagged in this step are recorded into the flag table FG3.

\subsubsection{Digital sampling correction}\label{sec:digitsampl}

Recording VLBI data involves digitizing the analog signals received at each antenna. However, real digitizers introduce gain errors due to nonideal sampling. These deviations can be estimated and corrected by analyzing the mean of the auto-correlation spectra, which should be normalized to unity. In \texttt{AIPS}, this correction is performed using the task \textit{ACCOR}, which computes the necessary scaling factors and stores them in the SN2 solution table. The corrections are then applied to the visibilities by the \textit{CLCAL} task, updating the calibration table version to CL5. This correction is always required for both the legacy and the new DiFX correlators at the VLBA. For some correlators, where the correction is applied prior to data output, reapplying it in \texttt{AIPS} is harmless and does not affect the data.

\subsubsection{Instrumental delay}

Instrumental delays arise from time and phase offsets introduced by hardware components such as cable mismatches, local oscillator discrepancies, and baseband converter delays. These effects typically produce stable, IF-dependent phase offsets that do not vary significantly over time. Correction involves fringe fitting on a strong calibrator to determine these offsets, which are then extrapolated to the rest of the scans to remove any IF dependent effects.

In \texttt{VIPCALs}, an antenna-based strategy is adopted. Using the calibrator scans identified for each antenna in Section~\ref{sec:calibchoice}, fringe fitting is performed on the central portion of each scan. The derived solutions are then extrapolated to the rest of the data. Although both delay and rate solutions are computed during this step, the rates are subsequently zeroed. This is because rate errors, which are time-dependent phase variations, are typically the result from atmospheric fluctuations or inaccuracies in the assumed source coordinates, and do not carry over between sources. Despite rates not being applied, solving for them improves the initial coarse search estimate from \textit{FRING}, which is later reflected in the fringe S/N. This approach ensures that instrumental effects are corrected using the brightest calibrator observed by each antenna. The solutions are stored in the SN3 solution table, and applied to the data via the CL6 calibration table.
It is worth noting that VLBA observations often include phase-calibration (PC) tables designed to track instrumental delays, albeit with varying success. \texttt{VIPCALs} adopts a more general approach that does not rely on the presence of such tables, which are sometimes absent or unreliable.

\subsubsection{Complex bandpass}\label{sec:bpass}
In addition to instrumental delays, another key instrumental effect is the bandpass response. Due to imperfections in receiver filters, their frequency response deviates from an ideal rectangular shape. Correcting for this distortion is essential before any averaging is performed. \texttt{AIPS} can address this with the \textit{BPASS} task, which derives a complex bandpass calibration. It uses the cross-power spectra, which are baseline-based, and solves for the antenna-based response via a least-squares fit. 
As in the previous calibration step, \texttt{VIPCALs} uses the selected calibrator scans for each antenna to solve for the bandpass. Because calibrator scans may vary across antennas and sources may have different flux densities, the parameter 
\textit{bpassprm(5)} is set to zero. This normalizes each time record by “channel zero”, which is by default defined as the central 75\% of channels in each IF. If the same calibrator scan is selected for all antennas, \textit{bpassprm(10)} = 6 is also set to shift the average bandpass phase to zero. This option should only be applied when calibration phases are stable, which may not hold if multiple sources are used for different antennas. On VLBA data, the \textit{BPASS} task also flags the outermost channels of each IF. This is caused by a small frequency shift due to the relative motions of the antennas with respect to the center of the Earth, which is used as the reference frame in the correlator. This becomes more noticeable in observations with few channels. All these solutions are stored in the bandpass table BP1.

Following bandpass calibration, the task \textit{ACSCL} is run to renormalize the amplitude scale based on autocorrelations, correcting any residual amplitude differences introduced by the bandpass correction. This produces the solution table SN4, which is applied to the data in CL7.

\subsubsection{Amplitude calibration}
Amplitude calibration in VLBI involves converting the raw correlator output into physical flux density units (Janskys, Jy). This is achieved by using the system equivalent flux density (SEFD) for each antenna, which accounts for the system temperature and the antenna gain. The SEFD of each antenna is calculated as

\begin{equation}
    \text{SEFD} = \frac{T_{\text{sys}}}{\text{DPFU} \cdot \text{gc}} 
\end{equation}

where $T_{\text{sys}}$ is the system temperature in Kelvin, DPFU is the antenna gain in Kelvin per Jansky (K/Jy), and $\text{gc}$ is the gain correction factor, derived from gain curves. Then, for a visibility amplitude \( A_{ij} \) measured on a baseline formed by antennas \( i \) and \( j \), the corresponding correlated flux density, \( S_{ij} \) (in Jy), is given by

\begin{equation}
    S_{ij} = A_{ij} \cdot \frac{\sqrt{\text{SEFD}_i\cdot\text{SEFD}_j}}{\eta}
\end{equation}

where $\eta$ is the quantization efficiency of the digital sampler, which has already been accounted for in \ref{sec:digitsampl}.

In \texttt{AIPS}, this calibration is performed using the task \textit{APCAL}. The calibration solutions produced by \textit{APCAL} are stored in the solution table SN5 and are subsequently applied to the calibration table version CL8. 

It is important to note that opacity corrections are not applied at this stage in the current version of the pipeline. Applying such a correction in \texttt{AIPS} requires either estimates of the atmospheric opacity from external measurements, or observations of a calibrator with a well-established flux density to derive the correction empirically. This limitation affects the accuracy of amplitude calibration at higher frequencies (e.g., above $\sim$22\,GHz), where atmospheric opacity becomes nonnegligible due to increased absorption by water vapor and other atmospheric constituents \citep{Carilli1999}. Alternative approaches exist for estimating the atmospheric opacity using the attenuated brightness temperature of the atmosphere in combination with system temperature measurements \citep{Janssen2019}.  However, the accuracy of these methods depends heavily on the sampling of the system temperatures at different airmasses; short observations or those with sparse measurements can lead to poor opacity fits and unreliable corrections. For this reason, opacity corrections have not yet been implemented in the current pipeline and are left for future development.

\subsubsection{Target fringe fit}\label{sec:tffit}
The final calibration step involves fringe fitting the science targets to correct for any remaining residual phase, delay, and fringe rate errors. This is once more performed using the task \textit{FRING}. The optimal solution interval is determined empirically by running a preliminary \textit{FRING} over a short segment of data and testing various solution intervals. The goal is to find the shortest interval that allows for the detection of fringes with a fringe S/N $\geq 5$ on all baselines. If such an interval cannot be found, the entire scan duration is used. The user can define minimum and maximum solution intervals to constrain the search. For the SMILE project, which primarily involves centimeter-wavelength data, the default minimum and maximum solution intervals are set to 1 minute and 10 minutes, respectively. These defaults have been chosen based on typical atmospheric coherence times at these wavelengths.

Once an optimal solution interval is set, the fringe fit is repeated to solve for single-band delay and fringe rate. During this step, the pipeline uses the \textit{aparm(9)} and \textit{search} options in \textit{FRING}. As described in \ref{sec:fringefitAIPS}, these options allow the task to search alternative antenna combinations in case of fringe detection failure on certain baselines. The priority of antenna selection follows the reference antenna strategy described in Section~\ref{sec:refant}. The rest of the parameters are 1000 ns and 200 mHz for the initial delay and rate windows, and a fringe S/N threshold of 5.  If the single-band fringe fit fails to produce reliable solutions across all baselines and time intervals, the pipeline attempts a multi-band delay solution using all IFs together by setting \textit{aparm(5)} = 1. This improves the fringe S/N by coherently combining signal across all IFs, assuming that residual single-band delays and phase-offsets have been adequately corrected in prior steps. The pipeline then selects the solution (single-band or multi-band) that yields the highest success rate, defined as the ratio of valid solutions to the total number of expected solutions. These fringe solutions are stored in the solution table version SN6 and applied to the calibration table CL9 using the \textit{CLCAL} task.

The pipeline also supports phase-referencing observations \citep{Beasley1995}, although it does not detect phase calibrators automatically. If the user provides a source–phase calibrator pair, the pipeline will perform fringe fitting on the phase calibrator using the same procedure described above. The resulting solutions are then transferred to the target source by interpolation using the \textit{CLCAL} task with the \textit{2pt} interpolation method.

\subsection{Data export and outputs}\label{sec:outputs}
Once the residual phases, delays, and rates have been corrected via fringe fitting or phase-referencing, and as long as valid solutions were found for at least one baseline, the calibrated data is split and exported. Splitting is performed using the \texttt{AIPS} task \textit{SPLIT}, which applies the final calibration table (CL9), the bandpass correction, and the flags to the visibility data. The data are then exported using the task \textit{FITTP}. By default, the pipeline averages all frequency channels within each IF, producing a single visibility point per IF. This averaging behavior can be modified by the user to preserve channels if needed. The exported visibilities are written in the \textit{uvfits} format,\footnote{\url{https://www.aips.nrao.edu/TEXT/PUBL/AIPSMEM117.PDF}} and can be subsequently loaded and imaged using other software packages such as \texttt{CASA}, \texttt{Difmap} \citep{Shepherd1997}, or \texttt{eht-imaging} \citep{Chael2019}. All calibration tables are also exported in a separate \textit{uvfits} file, allowing for re-import within \texttt{AIPS} if necessary.

\begin{figure}[h]
    \centering
    \includegraphics[width=0.47\textwidth]{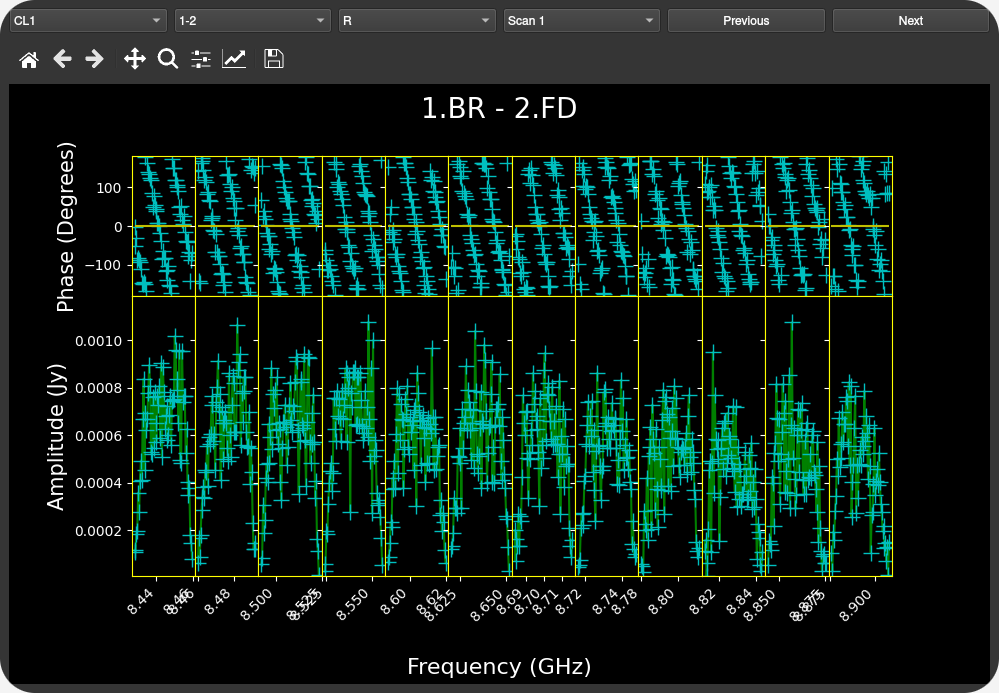}\\[1ex]
    \includegraphics[width=0.47\textwidth]{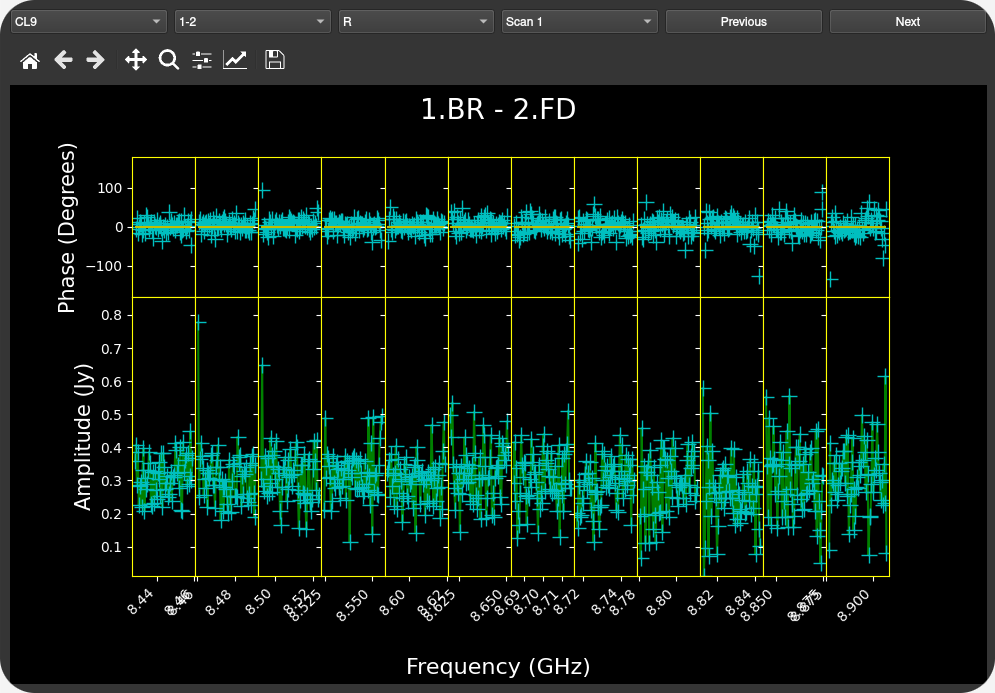}
\caption{Example of a single baseline before and after calibration, as displayed in the \texttt{VIPCALs} GUI. The plots show amplitude and phase as a function of frequency. The top figure corresponds to uncalibrated data, while the bottom figure shows the same baseline after applying the full calibration pipeline. Calibration flattens the phase response across the band and properly scales the amplitudes, correcting also for band-dependent variations.}

    \label{fig:uncalcalexample}
\end{figure}

To facilitate visual inspection of the calibration results, the pipeline produces a suite of diagnostic plots. These include
\begin{itemize}
    \item Uncalibrated and calibrated amplitude and phase versus frequency plots, using \textit{POSSM};
    \item Calibrated amplitude and phase versus time plots, using \textit{VPLOT};
    \item uv-plane coverage plots, using \textit{UVPLT};
    \item Visibility amplitude as a function of uv-distance.
\end{itemize}
If the user selects the “interactive” option in the GUI, these plots are also displayed using \texttt{matplotlib}, which allows for interactive inspection of how each calibration table affects the visibility data (see Figure \ref{fig:uncalcalexample} for an example). Additional examples of pipeline outputs, diagnostic plots, and the GUI interface are available in the pipeline documentation.\footnote{\url{https://vipcals.readthedocs.io}}

All text outputs from \texttt{AIPS} tasks are recorded and printed to a file. Additionally, a human-readable summary log is generated to track the outcome of each pipeline step. This summary is displayed in the GUI and also saved to disk. Finally, a comprehensive CSV file is produced containing metadata from the pipeline run. This file includes timing information for each step, the number of visibilities processed, antenna and calibrator scan rankings, flagged system temperature records, and other diagnostic metrics.

\section{Test on VLBA sample}\label{sec:sample}
The primary objective of the pipeline is to perform automated calibration of VLBA data for the SMILE project. The SMILE sample, defined in \cite{Casadio2021}, comprises 4968 radio sources selected from the sample in the VLA Cosmic Lens All Sky Survey (CLASS, \citealt{Myers2003}). A flux density threshold of 50 mJy at 8.4\,GHz is applied to ensure that each source is suitable for fringe fitting.

To select the data for SMILE, we searched the VLBA archive for observations corresponding to each of the 4968 sources using the following selection criteria:
\begin{itemize}
    \item Observations must include at least C-band or X-band  data.
    \item The correlation phase center must lie within 1 arcminute of the source coordinates reported in CLASS.
    \item Multiple files may be associated with a single observation of a source, provided they belong to the same VLBA project, share an identical frequency setup, and were observed within a two-day interval.
\end{itemize}

When multiple archival observations satisfied the above criteria for a given source, we selected the dataset with the longest on-source integration time. For sources with no matching archival data, new VLBA observations were obtained (BP267, PI: Pötzl; 20h).

To evaluate the performance and robustness of the pipeline, we constructed a representative subsample of 1000 sources based on the following stratified selection: 450 sources uniformly distributed in flux density between 50 mJy and 1 Jy; 450 sources uniformly distributed in total exposure time, ranging from 1 minute to 1 hour; and 100 sources that include observations at higher frequencies (U-band and K-band).

\begin{figure*}[t] 
  \centering
  \setlength{\tabcolsep}{0pt} 
  \begin{tabular}{@{}c@{\hspace{15pt}\vrule\hspace{15pt}}c@{}}
    \includegraphics[width=0.46\textwidth]{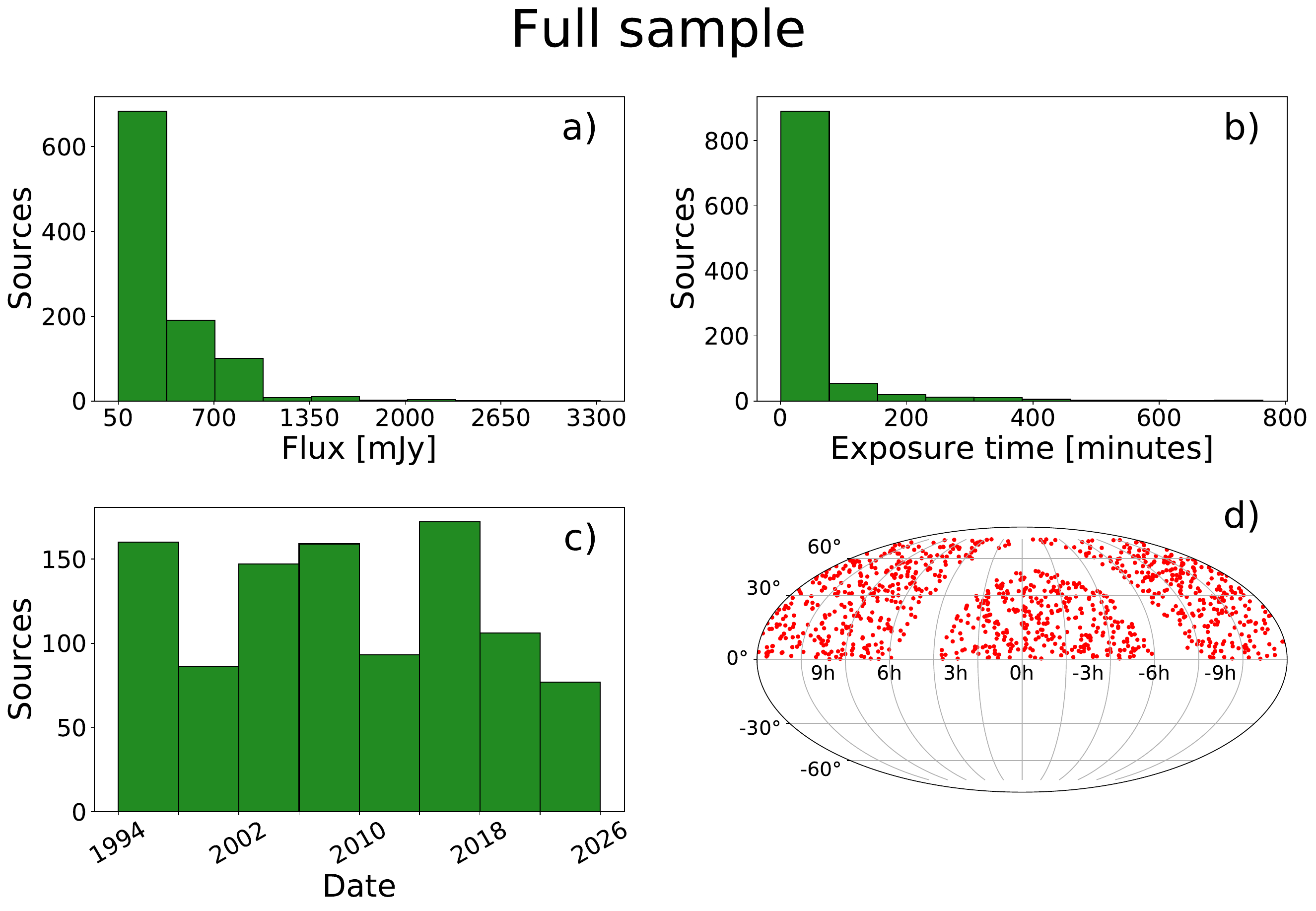} &
    \includegraphics[width=0.46\textwidth]{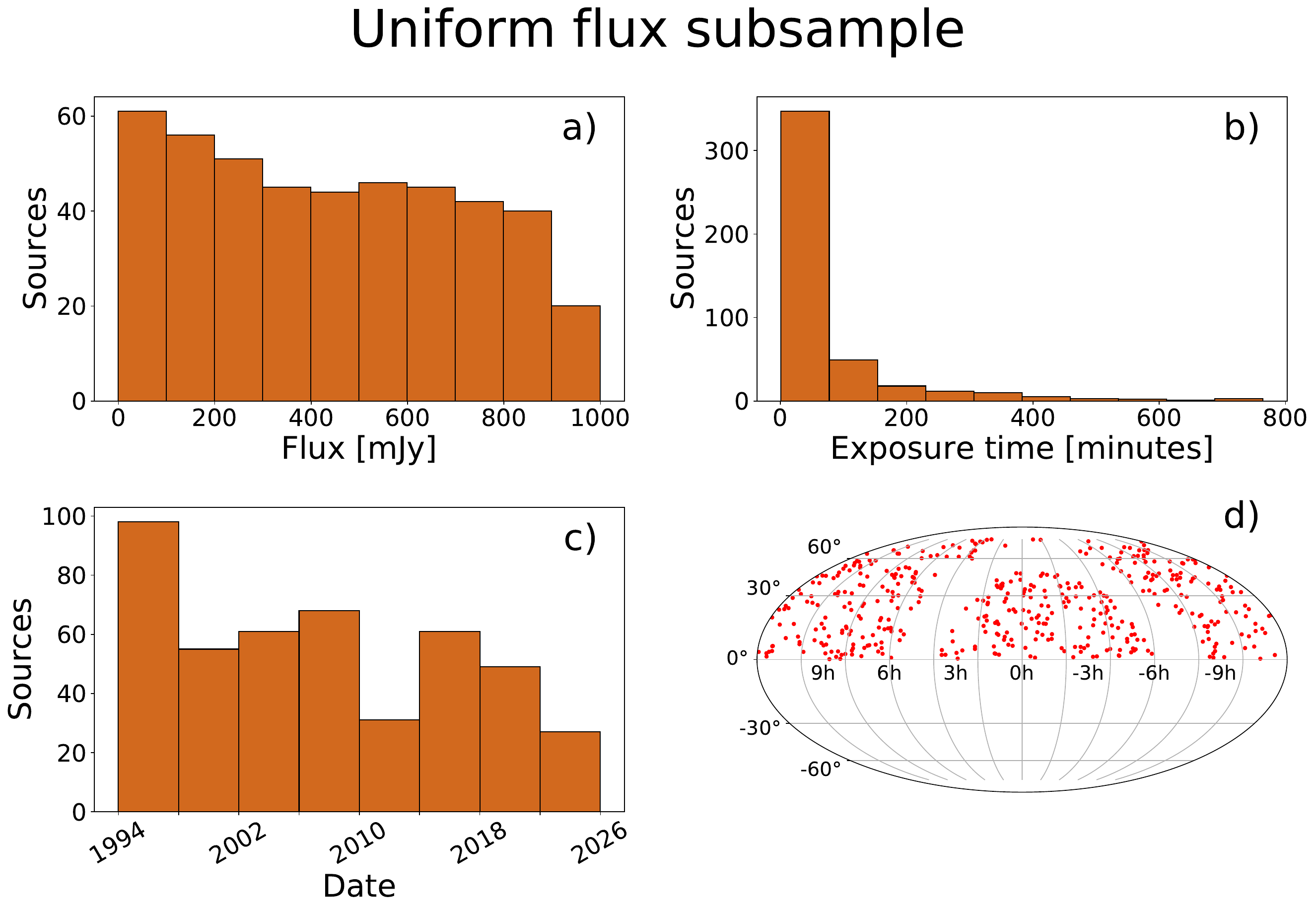}  \\ \hline
    \\ 
    \includegraphics[width=0.46\textwidth]{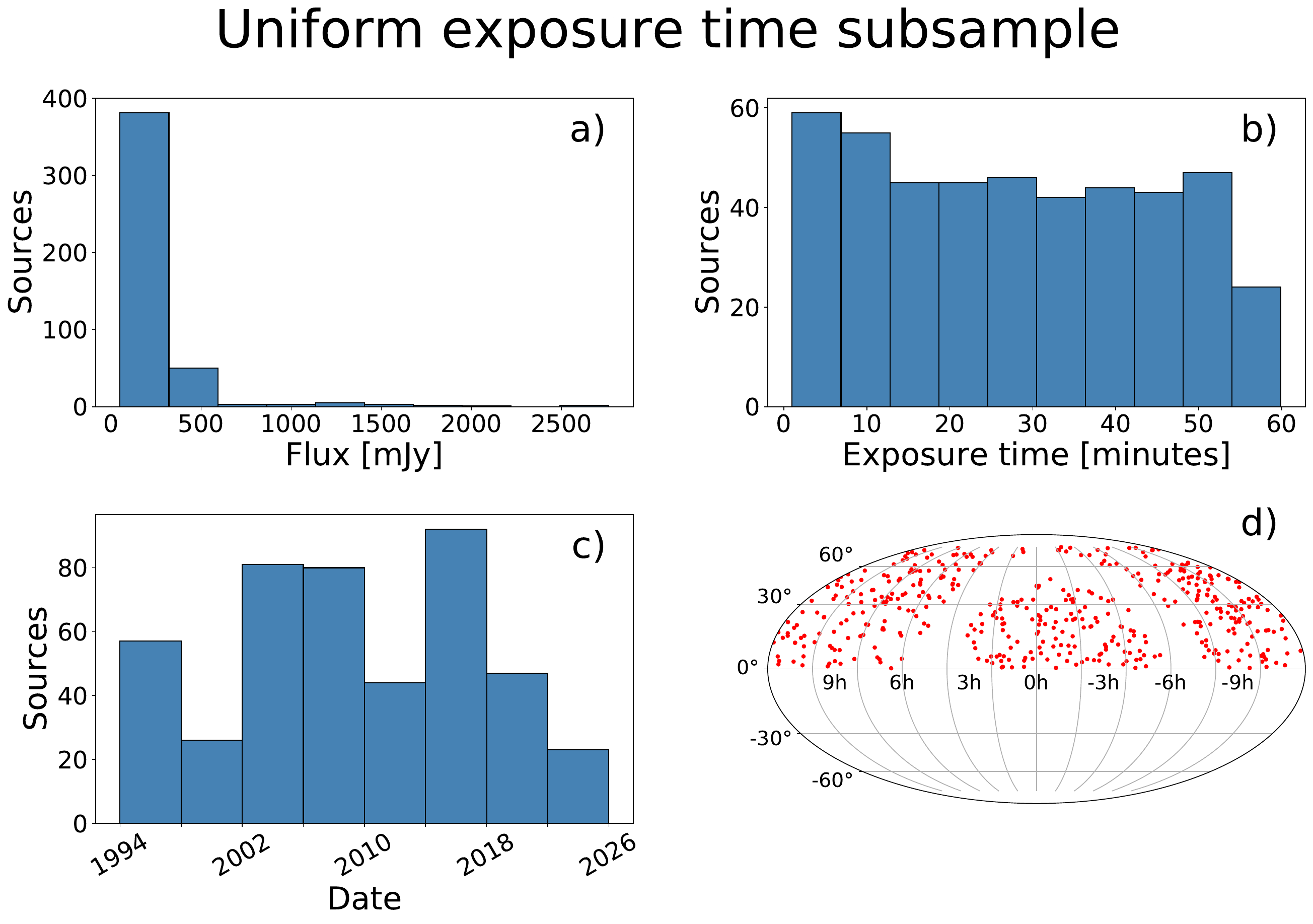} &
    \includegraphics[width=0.46\textwidth]{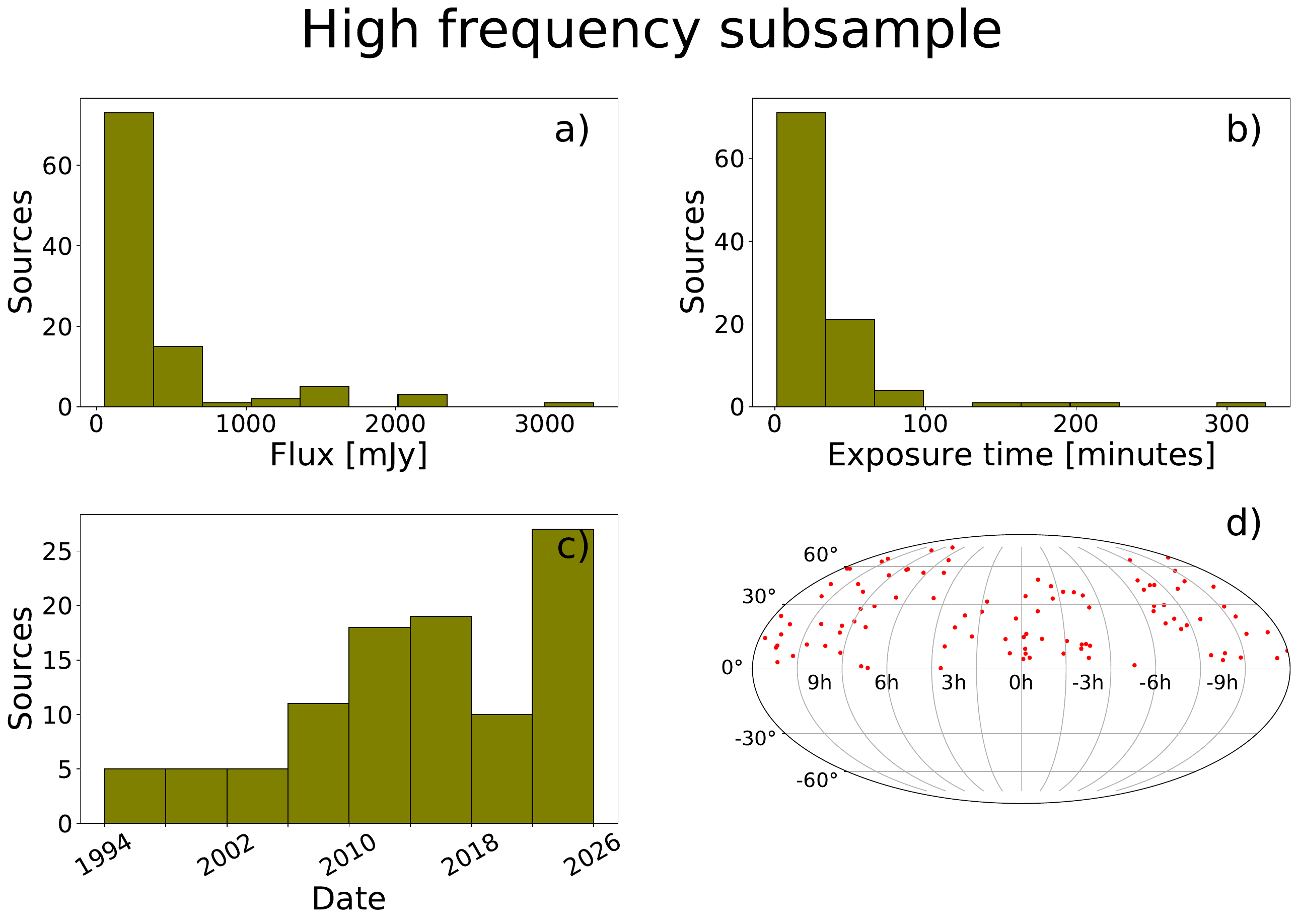}
  \end{tabular}
    \caption{
    Properties of the VLBA test sample used in this study. Each panel corresponds to a different subsample:
    \textit{Top left}: Full sample of 1000 sources.
    \textit{Top right}: Subsample uniformly distributed in (VLA) flux (450 sources).
    \textit{Bottom left}: Subsample uniformly distributed in exposure time (450 sources).
    \textit{Bottom right}: Subsample randomly selected from observations at 15 and 22\,GHz (100 sources).
    Each panel contains four subplots:
    (a) Histogram of source fluxes at 8.4\,GHz from VLA observations of the CLASS survey \citep{Myers2003}.
    (b) Histogram of on-target exposure times.
    (c) Distribution of observation dates.
    (d) Sky distribution in equatorial coordinates.
    }
    \label{fig:sample_properties}
\end{figure*}

The flux density values refer to VLA measurements at 8.4\,GHz from the CLASS survey, which served as the selection basis for the SMILE sample. While these flux densities do not directly represent the correlated VLBI flux (due to differences in sampled spatial frequencies) they remain a useful proxy for distinguishing between relatively bright and faint sources. The exposure times were obtained from the metadata of the corresponding VLBA observations.

Although SMILE requires at least C or X band data, many of the selected archival datasets include additional frequency bands, all of which were processed through the pipeline. If we define an observation as a unique source–frequency pair, the test sample includes 1417 observations of 1000 sources, drawn from 360 different VLBA projects. These datasets are distributed across 2589 files and occupy a total data volume of 19~TB. The data cover also more than 3 decades of VLBA observations, spanning from 1994 up to 2025. An overview of the test dataset properties is shown in Figure \ref{fig:sample_properties}. This comprehensive test sample provides a robust benchmark for evaluating the pipeline’s calibration accuracy, performance under varying S/N conditions, and ability to handle heterogeneous legacy VLBA datasets.

\section{Results and discussion}\label{sec:results}

The pipeline successfully completed all processing steps for 955 sources, which correspond to 1372 individual observations. For the remaining 45 sources, the calibration could not be finalized, and the pipeline halted at an intermediate stage. Section~\ref{sec:calibtime} presents a breakdown of the time required by the pipeline. Section~\ref{sec:dataevolution} examines how the data evolved throughout the calibration process. The results of the fringe-fitting procedure on the science targets are discussed in Section~\ref{sec:ffitsolutions}. Finally, in Section~\ref{sec:faileddata} we analyze the characteristics of datasets that failed to calibrate successfully.

\subsection{Calibration time}\label{sec:calibtime}
The test was conducted on a system equipped with an Intel\textsuperscript{\textregistered} Xeon\textsuperscript{\textregistered} Silver 4314 CPU (16 cores, 2.4\,GHz base frequency). The observational data were stored on Western Digital Ultrastar DC HC560 20\,TB hard disk drives, each connected via SATA III and with a sustained read speed of 291\,MB/s. The processing in \texttt{AIPS} was done on a Dell Enterprise NVMe AGN MU 1.6\,TB SSD, with read and write speeds of up to 6.2\,GB/s and 2.8\,GB/s respectively. The SSD was mounted on four PCIe Gen 4.0 lanes. Although \texttt{ParselTongue} permits trivial parallelization, i.e., launching independent \texttt{AIPS} tasks simultaneously, the pipeline was executed in single-core mode. This decision was motivated by the relative simplicity of implementation, combined with satisfactory performance in practice. Since \texttt{AIPS} remains constrained to a maximum of 12\,GB of RAM in a 64-bit system, memory was not a limiting factor. Instead, profiling indicated that disk input/output (I/O) speed was the primary bottleneck, further justifying the single-threaded execution model.

\begin{figure}[t]
    \centering
    \includegraphics[width=0.47\textwidth]{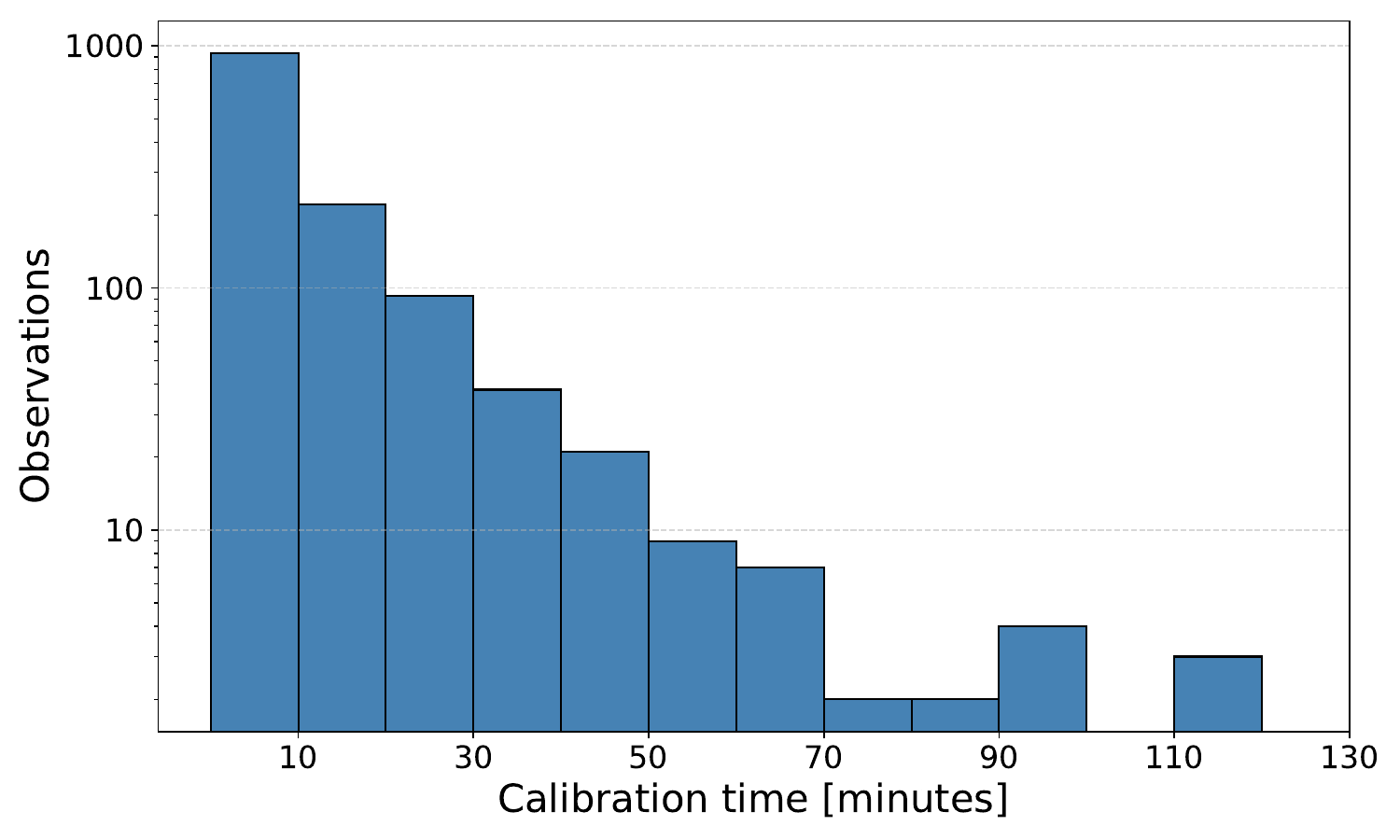}
    \caption{Time needed to calibrate each of the 1372 VLBA observations using \texttt{VIPCALs}. Mean and median run times were 9.5 and 4.2 minutes, respectively.}
    \label{fig:timehistogram}
\end{figure}

\begin{figure}[t]
    \centering
    \includegraphics[width=0.47\textwidth]{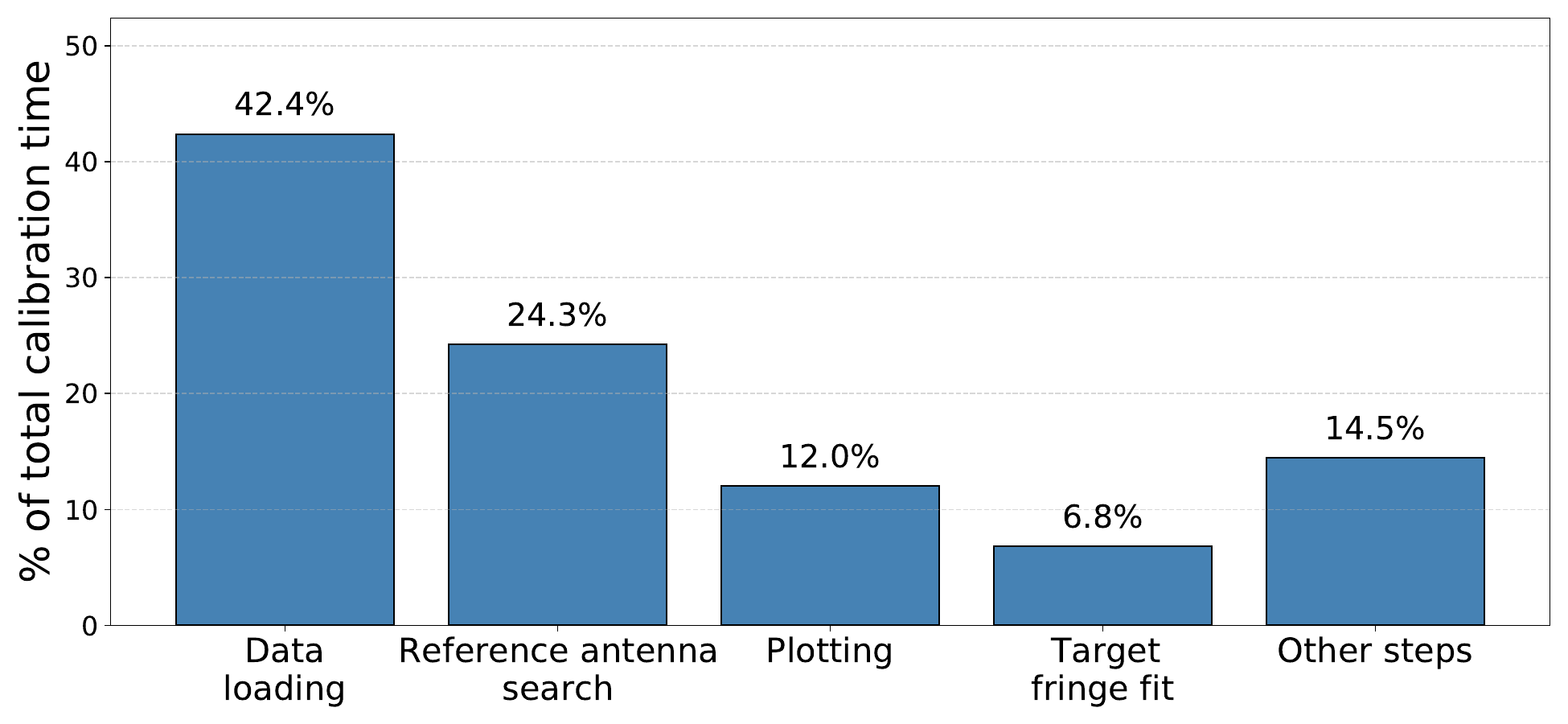}
    \caption{Breakdown of the relative time spent across different pipeline steps for the 1372 VLBA observations calibrated with \texttt{VIPCALs}. More than 50\% of the calibration time is spent in I/O intensive steps (data loading and plotting).}
    \label{fig:timepiechart}
\end{figure}

The complete dataset of 1372 observations was calibrated in approximately 7.5 days. On average, the calibration of each source–band pair required around 9.5 minutes, with run times ranging from 16 seconds (minimum) to 2 hours (maximum). The median calibration time was 4.2 minutes per observation. The distribution of the run times is shown in Figure~\ref{fig:timehistogram}. These processing times represent a substantial improvement over manual calibration workflows, where even a basic visual inspection of the data often exceeds this duration. Figure~\ref{fig:timepiechart} presents the relative share of total calibration time consumed by some of the pipeline stages. The most time-consuming components are those associated with intensive read/write operations, i.e., reading and loading data into \texttt{AIPS}, and generating and plotting the different diagnostic plots. Overall, the results support the suitability of the pipeline for scalable and automatic VLBA data processing. While limited by disk I/O throughput, the current configuration already represents a significant improvement in efficiency and reproducibility over traditional manual calibration approaches.

\subsection{Visibility count across calibration}\label{sec:dataevolution}
\begin{figure*}[h!]
    \centering
    \includegraphics[width=0.95\textwidth]{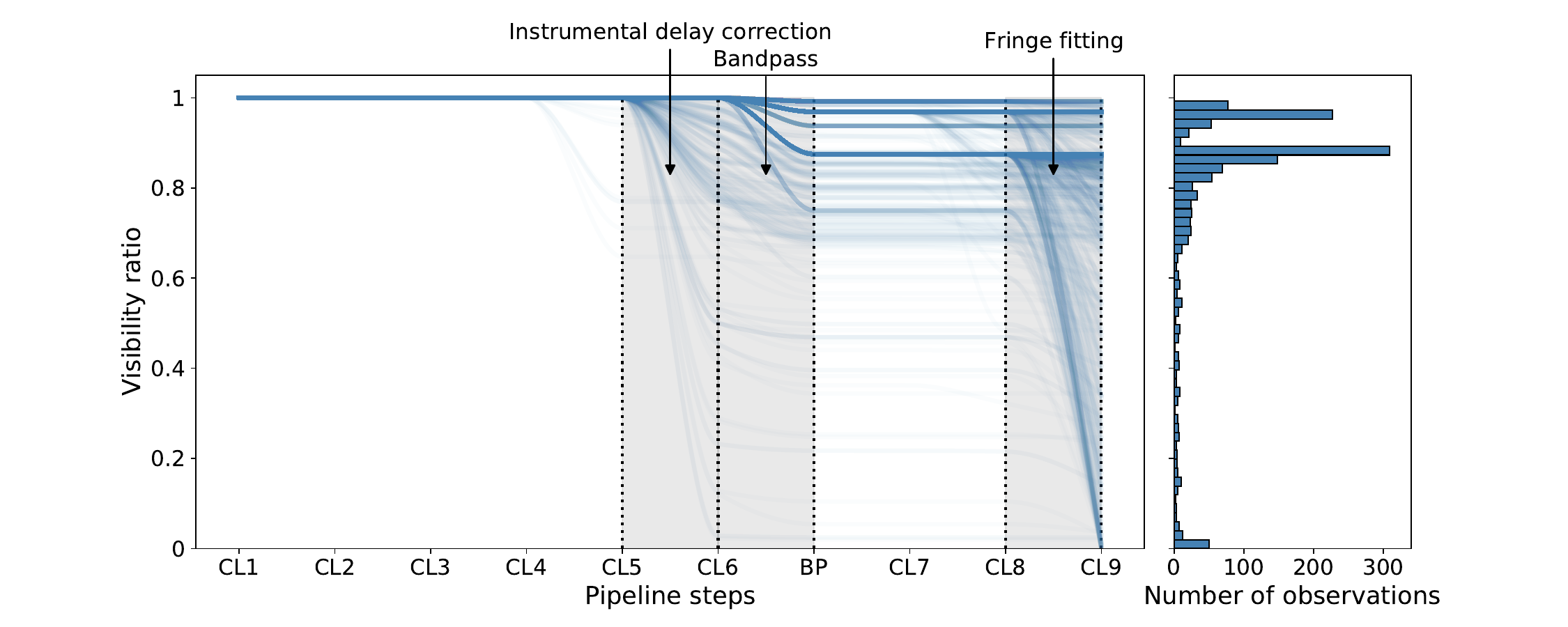}
    \caption{Visibility ratio across the calibration pipeline for 1372  VLBA observations. The left panel shows, for each observation, the visibility ratio at each major calibration stage. Each semi-transparent line represents a single observation, so darker regions correspond to a higher density of observations. The most significant drops are highlighted in gray. They correspond to the instrumental delay correction (CL5 to CL6), bandpass calibration, and the final fringe fitting on science targets (CL8 to CL9). The right panel shows the distribution of final visibility ratios after the full pipeline.}
    \label{fig:calibtimeline}
\end{figure*}

To assess pipeline performance, we track the number of initial raw visibilities and the number of visibilities retained after each calibration stage, i.e., after the application of each calibration table. For that purpose we define the visibility ratio as the ratio between unflagged and total visibilities at each step. Figure~\ref{fig:calibtimeline} illustrates the visibility ratio across the pipeline for all 1372 successfully calibrated observations.

The left panel of Figure~\ref{fig:calibtimeline} shows, for each observation, the ratio of visibilities retained at each calibration stage. Note that the initial visibility count excludes non-VLBA antennas for which system temperature or gain curve information was unavailable, and which were therefore omitted from the calibration process. The three calibration steps which account for most of the visibility losses are: (i) the instrumental delay calibration (CL5 to CL6), (ii) the application of the bandpass correction, and (iii) the final fringe fitting on the science target (CL8 to CL9).

The drop in visibilities following instrumental delay correction typically reflects datasets in which one or more antennas failed to detect strong fringes on the calibrator. This may be due to poor observing conditions on a specific antenna (e.g., bad weather) or insufficient calibrator strength, particularly on long baselines. The visibility loss after bandpass calibration stems from a small frequency shift introduced by the correction (see Section~\ref{sec:bpass}). This shift leads to the automatic flagging of edge channels in each IF. The extent of visibility loss varies with the spectral resolution: from under 1\% in datasets with 256 channels per IF, to as much as 25\% in cases with only 8 channels per IF. Nevertheless, this loss is not considered critical, as the S/N is known to degrade significantly at the edge channels of each subband. These edge channels are typically flagged in standard VLBA data reduction workflows due to digital filter roll-off and bandpass shape effects. The final visibility drop (CL8 to CL9) reflects cases where the fringe-fitting step on the science target failed to detect reliable residual delays or rates. Further discussion of this results is provided in Section~\ref{sec:ffitsolutions}.

The right panel of Figure~\ref{fig:calibtimeline} shows the final distribution of the visibility ratio after the full pipeline. This distribution serves as an approximate proxy for the overall calibration quality. We observe two prominent peaks in the distribution: one near a visibility ratio of 0.97 (typically datasets with 64 channels per IF and only two flagged channels due to bandpass corrections), and another around 0.88 (for 16-channel IFs). The median and mean visibility ratios across the full sample are 0.87 and 0.78, respectively. Overall, 87.4\% of the observations retain more than half of their original visibilities, while only 76 sources (7.2\% of the observations) retain fewer than 20\%. 

\subsection{Fringe fit solutions}\label{sec:ffitsolutions}
As shown in Figure~\ref{fig:calibtimeline}, the final fringe fit for residual delays and rates on the science targets is the calibration stage that removes the largest fraction of visibilities. This is expected: the \textit{FRING} task rejects solutions in which no fringe is detected above the adopted fringe S/N threshold of 5. We therefore define the fringe ratio for an observation as the fraction of the total fringe solutions that exceed this threshold. Since failed fringe solutions often reflect intrinsic source weakness, source structure on long VLBA baselines, or short on-source time, the fringe ratio serves as a practical proxy for data quality.

\begin{figure}[h]
    \centering
    \includegraphics[width=0.47\textwidth]{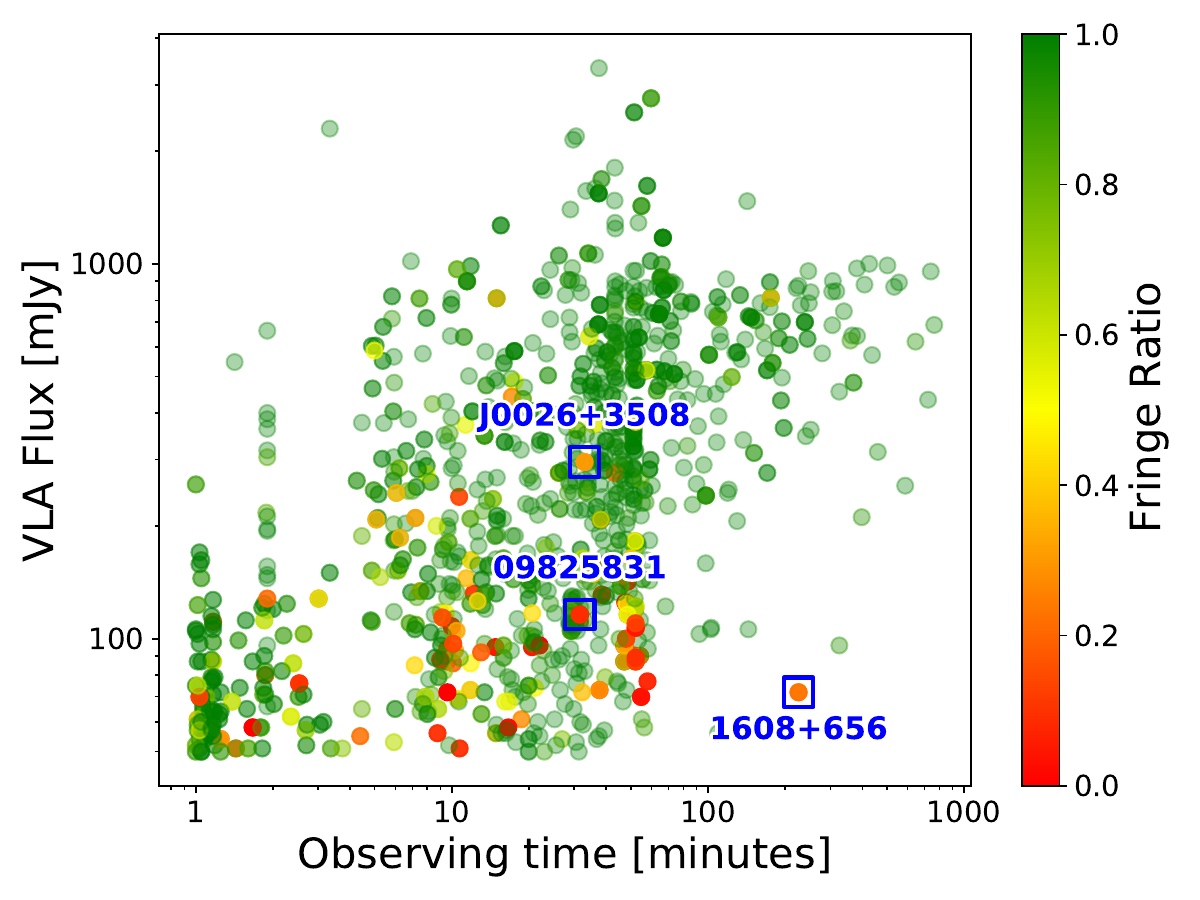}
    \caption{Ratio of successful fringe solutions as a function of reported VLA flux at 8.4\,GHz and exposure time for the 1372 VLBA observations calibrated using \texttt{VIPCALs}. Blue squares highlight three sources with low fringe ratio despite having a high flux and/or observation time. These examples are further discussed in the main text.}
    \label{fig:ffitcolor}
\end{figure}

To explore how detectability depends on source strength and exposure time, Figure~\ref{fig:ffitcolor} plots the fringe ratio for different datasets as a function of the source's reported VLA flux and exposure time. The majority of observations (91.6\%) achieve fringe detections in at least half of the attempted solution intervals, indicating that the adopted workflow is adequate for most of the sample. However, a small number of observations with high flux and/or observation time fall well below this trend. Some examples are

\begin{itemize}
    \item J0026+3508: Fringe ratio 0.30 - C band, BC151 (2005).  
    \item 09825831: Fringe ratio 0.09 - C band, BF072 (2004). 
    \item 1608+656: Fringe ratio 0.23 - X band, BF067 (2001). 
\end{itemize}

In these cases, it is helpful to compare the reported 8.4\,GHz VLA flux densities and the actual correlated flux on VLBA baselines. For example, VLBA observations of J0026+3508 from the Astrogeo VLBI image database\footnote{\url{http://astrogeo.org/vlbi_images/}} \citep{Petrov2025} show a total flux density of 100 mJy in C band, with a peak flux of 40 mJy/beam, compared to the total flux of 296 mJy measured by the VLA. A similar case is 09825831, which has a correlated VLBA flux of 37 mJy (J09496+5819 in \citealt{Helmboldt2007}) versus 116 mJy reported by CLASS. For 1608+656, no VLBI images are available, but the flux on the shortest VLBA baselines is estimated at 35 mJy, compared to 72 mJy from CLASS. These differences show how, in some cases, the VLA flux densities can substantially overestimate the correlated flux on VLBA scales, and that poor fringe-fitting performance could be a consequence of intrinsic weakness.

In all three examples, manual inspection of intermediate calibration products indicates that earlier pipeline choices (reference antenna, calibrator scans, solution intervals) are adequate. Averaging polarizations before fringe fitting (\textit{aparm(3)} = 1 in \textit{FRING}) improves the fringe ratio for 1608+656 from 0.23 to 0.50, but has negligible effect in the other cases. Examining other datasets with low fringe ratios yields similar conclusions: limited fringe detections are primarily driven by astrophysical or observational factors rather than shortcomings in the pipeline. In such cases, lowering the S/N threshold in \textit{FRING} might recover additional detections, but at increased false-positive risk. As the goal of \texttt{VIPCALs} is to provide a robust unsupervised calibration, a default conservative threshold is preferred, although it can always be defined by the user. 

Overall, the fringe-fitting performance of \texttt{VIPCALs} is robust across the majority of the test sample, with most observations returning reliable solutions despite the heterogeneous nature of the archival data. Cases with poor fringe recovery are associated with more challenging datasets and do not seem to result from incorrect pipeline decisions. Improving the calibration of these cases would require manual inspection, fine tuning of fringe-fitting parameters, or incorporation of external source models, all of which demand significant time and expertise. As such, the conservative and fully automated approach adopted by \texttt{VIPCALs} provides a practical balance between reliability and completeness, with the understanding that a small fraction of difficult datasets will fall outside the reach of unsupervised calibration.

\subsection{Noncalibrated datasets}\label{sec:faileddata}
Out of the 1000 sources selected for the test sample, the pipeline was unable to fully calibrate the datasets corresponding to 45 sources. A detailed inspection revealed that these failures arise from a limited set of issues, and they are summarized below.

Nonstandard system temperature tables caused failures for 20 sources. When system temperature (TY) and gain curve (GC) information is missing from the dataset, \texttt{VIPCALs} attempts to retrieve and format them from an online repository. However, several of these auxiliary tables do not follow standard formatting conventions, contain typographical errors, or include measurements in frequency bands unrelated to the actual observation. In some cases, the tables are simply not available. Since \texttt{AIPS} is sensitive to inconsistencies in these tables, and given the risk of introducing an incorrect amplitude calibration, the pipeline is intentionally conservative. No automated correction is attempted in these cases. Users may manually provide corrected TY and GC tables if desired.

Non-ordered IF setups resulted in failures for 13 sources. The pipeline infers different observing bands by grouping IFs based on their central frequencies. This process assumes that IFs belonging to a given band are consecutively numbered. However, some legacy VLBA observations (particularly from the 1994–1995 period) contain IFs with interleaved frequencies from multiple bands (e.g., 8 IFs with central frequencies 2.22, 2.23, 8.15, 8.23, 2.29, 2.32, 8.41, and 8.55\,GHz). Currently, the \texttt{AIPS} task \textit{FITLD} does not support loading non-consecutive IFs, which prevents the pipeline from initializing its workflow. Support for handling such configurations automatically is a work in progress.

The presence of subarrays gave rise to errors for ten sources. Despite applying a VLBA-only filter during data selection, mislabeling in the NRAO Archive led to the inclusion of some observations involving additional antennas (e.g., from global VLBI experiments). As a result, some datasets contain subarray configurations, where different groups of antennas observed different sources simultaneously. Proper handling of those datasets requires the \textit{USUBA} task in \texttt{AIPS}, which can assign subarray indices based on scan structure. While an automatic mode for \textit{USUBA} exists, it carries significant risk of misassignment and typically requires later verification, an approach incompatible with the goals of full automation. Although \texttt{VIPCALs} includes partial support for subarray detection, certain tasks may still fail due to inconsistent indexing when subarrays are not correctly defined. Manual subarray definition is planned for future versions.
    
Finally, incorrect ordering information in the header prevented successful calibration for 2 sources. \texttt{VIPCALs} uses metadata in the \texttt{FITS} file header to determine whether the dataset requires reordering into the standard time–baseline format. When this header information is incorrect, the data remain unordered, preventing \texttt{AIPS} from executing subsequent tasks. Although the pipeline could, in principle, verify or reorder the data by default, both operations are computationally expensive. Given the very low occurrence of such cases, the current approach is to rely on the header information and allow processing to fail when it is wrong. 

These failures  are concentrated in a relatively small number of archival projects and do not reflect systemic limitations of the pipeline. Most could be addressed through either minor user intervention or future extensions of the current workflow.

\section{Future work}\label{sec:futurework}
Beyond addressing the limitations discussed in Section~\ref{sec:faileddata}, ongoing and future developments of the \texttt{VIPCALs} pipeline will focus on expanding compatibility across broader classes of VLBI observations and improving user experience. One priority is the automated handling of non-VLBA antennas. While the pipeline currently retrieves and processes system temperature and gain curve data for VLBA antennas, many archival projects labeled as VLBA also include antennas from the High Sensitivity Array (HSA) (e.g., Green Bank Telescope, Effelsberg, or Arecibo). These are large and sensitive antennas, and their inclusion could significantly enhance the final data quality. Another key objective is extending pipeline compatibility to other VLBI networks, such as the EVN. While the calibration workflow of \texttt{VIPCALs} is expected to generalize well to any VLBI observation, specific adjustments are required to account for differences such as metadata conventions between arrays. Additionally, large arrays often include simultaneous or subarrayed scans, revisiting the need for more robust subarray handling within the pipeline framework.

An increased frequency coverage also constitutes a priority. Although \texttt{VIPCALs} has been extensively validated on cm-wavelength VLBA data, calibration at both lower and higher frequencies introduces unique challenges. At high frequencies (22\,GHz and above), atmospheric opacity must be accounted for during amplitude calibration. At low frequencies, steep source spectral indices across wide fractional bandwidths must be modeled or compensated for, and radio frequency interference (RFI) becomes increasingly problematic. Automated RFI flagging routines will be essential for robust low-frequency performance.

Another possible direction is extending the current workflow, designed for continuum, total-intensity observations, to polarization calibration. Full-Stokes processing requires several additional steps beyond the parallactic-angle correction already implemented. The cross-hand phase and delay offsets between the two polarization channels must first be determined, typically through fringe fitting on a strong polarized calibrator. Subsequently, polarization leakage between feeds (the so-called D-terms) must be estimated and corrected. While this is ideally done using an unpolarized calibrator, such sources are rarely available in practice. Instead, existing total-intensity models are often subdivided into regions of assumed constant fractional polarization, and the D-terms are solved by minimizing residuals with respect to these submodels. This method is implemented in the \texttt{AIPS} task \textit{LPCAL}, and in more recent tools such as \texttt{GPCAL} \citep{Park2021} and \texttt{polsolve} \citep{MartiVidal2021}. Automating this calibration sequence would require the pipeline not only to identify suitable polarization calibrators, but also to access reliable intensity models, and to assess the quality of the derived solutions, tasks that typically rely on human supervision in manual calibration workflows.

Similarly, supporting spectral-line observations would require substantial extensions to the calibration workflow. In this case, a strong continuum calibrator is used to obtain initial delay and rate solutions through fringe fitting. Because of the large number of spectral channels involved, the \textit{FRING} setup must be fine-tuned relative to the configuration described in Section~\ref{sec:tffit}. The strongest line(s) of the target source are then fringe-fitted to refine the delay-rate corrections. One of these lines must be defined as the reference spectral line, used to correct for Doppler shifts caused by the motion of the antennas relative to the source during the observation and across epochs. Bandpass solutions must also be examined carefully, as imperfections can introduce spectral artifacts, and additional steps are needed for accurate amplitude calibration. While full automation of all these steps remains challenging, a hybrid approach, where the user specifies spectral windows and lines of interest, and the pipeline performs all other tasks, could make \texttt{VIPCALs} suitable for maser and astrometric VLBI experiments.

Although further extensions of the workflow are planned, the present version of \texttt{VIPCALs} is already capable of large-scale applications. One particularly promising case is the systematic calibration of VLBA data in the NRAO Archive, which spans over three decades of observations across thousands of projects. Automating the calibration of archival data would provide the community with science-ready visibility datasets, lowering the barrier for researchers unfamiliar with VLBI calibration. This effort would not only enable historical observations to be used more easily but also unlock opportunities for large-scale statistical studies on the long-term behavior of radio-loud sources, instrumental stability, or observational strategies, among others.

Finally, as part of the science goals of the SMILE project, the next logical step is the development of an automated imaging solution. Initial testing of imaging algorithms is underway, and results will be reported in an upcoming publication.

\section{Summary and conclusions}\label{sec:conclusions}

We have presented \texttt{VIPCALs}, a fully automated calibration pipeline for continuum VLBI data developed in the context of the SMILE project but designed as a general-purpose tool. The pipeline interfaces with \texttt{AIPS} via \texttt{ParselTongue} and performs end-to-end unsupervised calibration of VLBI datasets, starting from raw visibility files and producing science-ready outputs. The \texttt{VIPCALs} workflow reproduces the standard \texttt{AIPS} calibration strategy, including precalibration processing, ionospheric and geometric corrections, instrumental delay and bandpass calibration, and a final fringe-fitting stage. The pipeline automatically handles tasks such as reference antenna selection, calibrator identification, and metadata retrieval while providing flexibility for more experienced users via a graphical interface and configuration options. The pipeline design prioritizes not only automation and scalability but also reproducibility, which ensures consistency in the calibration and facilitates its reanalysis.

We validated \texttt{VIPCALs} on a representative test sample of 1000 sources from the SMILE project covering a broad range of flux densities, integration times, and frequency setups in archival VLBA projects. The pipeline successfully calibrated 955 sources, with failures primarily due to nonstandard metadata or subarray configurations. Across the successfully calibrated observations, the median visibility retention ratio was 87\%, indicating minimal data loss throughout the pipeline. Additionally, over 91\% of the datasets achieved successful fringe fitting in at least half of the attempted solutions. On average, \texttt{VIPCALs} required just 9.5 minutes per observation to complete the full calibration workflow in single-threaded mode, significantly reducing the time and expertise traditionally required for VLBI data calibration. Most of the processing time was consumed by I/O-intensive steps, such as data loading and diagnostic plotting.

In summary, \texttt{VIPCALs} provides an efficient, scalable, and robust solution for calibrating large samples of VLBI datasets without user intervention. This capability is essential for enabling upcoming surveys that aim to process thousands of VLBI observations in a consistent and reproducible manner. In particular, \texttt{VIPCALs} has proven to be perfectly suited to the needs of the SMILE project, where the sheer volume and heterogeneity of the archival VLBA data make manual calibration unfeasible. Its ability to operate end-to-end without prior information while maintaining high calibration quality and diagnostic transparency, makes it an indispensable tool for the large-scale search of milli-lens candidates.

\begin{acknowledgements}
D.A., C.C., F.P. \& A.K. acknowledge support from the European Research Council (ERC) under the Horizon ERC Grants 2021 programme under grant agreement No. 101040021. The authors thank Dmitry Blinov and Loránt Sjouwerman for their valuable feedback on this work. We also thank Alejandro Mus for his assistance with the implementation and deployment of the pipeline. We acknowledge the staff at the NRAO’s Pete V. Domenici Science Operations Center (Socorro, NM) for their support in retrieving the archival data used in this study, and we extend our special thanks to Eric Greisen for his indispensable assistance and long-standing support of \texttt{AIPS}.\\
\end{acknowledgements}

\bibliographystyle{aa}
\bibliography{mybib.bib}

\begin{appendix}

\end{appendix}

\end{document}